\documentclass[sigconf]{acmart}

\AtBeginDocument{%
  }

\copyrightyear{2025}
\acmYear{2025}
\setcopyright{acmlicensed}\acmConference[UIST '25]{The 38th Annual ACM Symposium on User Interface Software and Technology}{September 28-October 1, 2025}{Busan, Republic of Korea}
\acmBooktitle{The 38th Annual ACM Symposium on User Interface Software and Technology (UIST '25), September 28-October 1, 2025, Busan, Republic of Korea}
\acmDOI{10.1145/3746059.3747709}
\acmISBN{979-8-4007-2037-6/2025/09}

\usepackage{enumitem}
\usepackage{wrapfig}
\usepackage{stfloats}
\usepackage[skip=4pt]{caption}

\newlength{\oldintextsep}
\definecolor{mypink}{RGB}{241, 225, 225}
\definecolor{myblue}{RGB}{204, 228, 240}

\newcommand{\add}[1]{{\color{black}{#1}}}

\newcommand{\CoolName}[1]{Reality Proxy}
\newcommand{\para}[1]{\vspace{-2mm}\paragraph{\textbf{#1}}}
\newcommand{\quot}[1]{\emph{``#1''}}

\makeatletter
\def\subsubsection{\@startsection{subsubsection}{3}%
  \z@{.5\linespacing\@plus.7\linespacing}{.1\linespacing}%
  {\normalfont\itshape}}
\makeatother

\begin{document}


\title[\CoolName{}: Fluid Interactions with Real-World Objects in MR]{\CoolName{}: 
Fluid Interactions with Real-World Objects in MR \\ via Abstract Representations
}

\author{Xiaoan Liu}
\authornote{This work was done during Xiaoan's summer internship at University of Minnesota.}
\email{xiaoanliu@nyu.edu}
\orcid{0009-0005-1294-7108}
\affiliation{%
  \institution{New York University}
  \city{New York}
  \state{New York}
  \country{USA}
}

\author{Difan Jia}
\email{difanjia@umn.edu}
\orcid{0009-0008-7470-5202}
\affiliation{%
  \institution{University of Minnesota}
  \city{Minneapolis}
  \state{Minnesota}
  \country{USA}
}

\author{Xianhao Carton Liu}
\email{liu03008@umn.edu}
\orcid{0009-0006-3528-9651}
\affiliation{%
  \institution{University of Minnesota}
  \city{Minneapolis}
  \state{Minnesota}
  \country{USA}
}

\author{Mar Gonzalez-Franco}
\email{margon@google.com}
\orcid{0000-0001-6165-4495}
\affiliation{%
  \institution{Google}
  \city{Seattle}
  \state{Washington}
  \country{USA}
}

\author{Chen Zhu-Tian}
\email{ztchen@umn.edu}
\orcid{0000-0002-2313-0612}
\affiliation{%
  \institution{University of Minnesota}
  \city{Minneapolis}
  \state{Minnesota}
  \country{USA}
}

\renewcommand{\shortauthors}{Liu et al.}

\begin{abstract}
Interacting with real-world objects in Mixed Reality (MR) often proves difficult when they are crowded, distant, or partially occluded, hindering straightforward selection and manipulation. We observe that these difficulties stem from performing interaction directly on physical objects, where input is tightly coupled to their physical constraints.
Our key insight is to decouple interaction from these constraints by introducing \emph{proxies}--abstract  representations of real-world objects. 
We embody this concept in \CoolName{}, 
a system that seamlessly shifts interaction targets from physical objects to their proxies during selection.
Beyond facilitating basic selection, 
\CoolName{} uses AI to enrich proxies with semantic attributes and hierarchical spatial relationships 
of their corresponding physical objects,
enabling novel and previously cumbersome interactions in MR-such as skimming, attribute-based filtering, navigating nested groups, and complex multi-object selections—all without requiring new gestures or menu systems.
We demonstrate \CoolName{}'s versatility across diverse scenarios, including office information retrieval, large-scale spatial navigation, and multi-drone control.
An expert evaluation suggests the system's utility and usability, suggesting that proxy-based abstractions offer a powerful and generalizable interaction paradigm for future MR systems.
\end{abstract}



\begin{teaserfigure}
  \includegraphics[width=\textwidth]{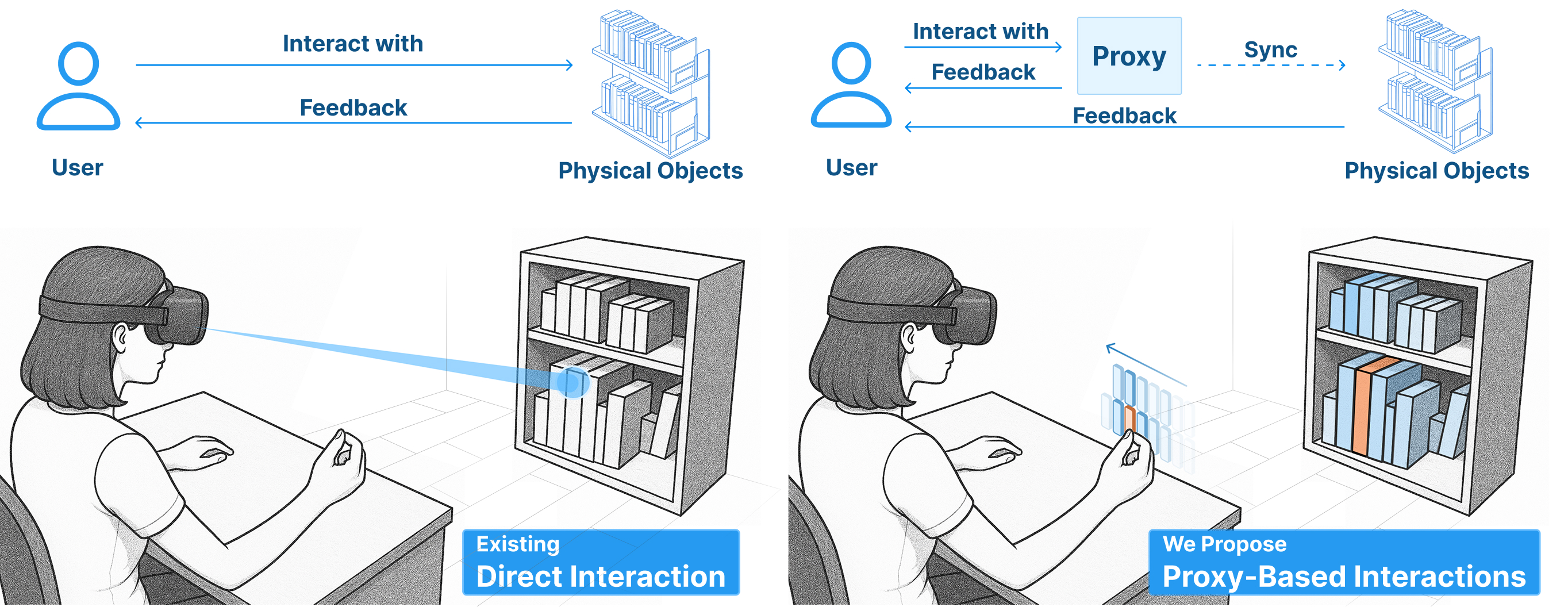}
  \caption{
    Existing direct interaction with physical objects (left) vs. our proxy-based approach (right). Rather than pointing at distant objects, \CoolName{} shifts the interaction target to a digital proxy synchronized with the real object, freeing the user from physical constraints like distance or size. 
    Enriched by AI-derived scene structures, these proxies also enable advanced interactions like multi-object selection, filtering, and grouping.
  }
  \label{fig:teaser}
\end{teaserfigure}


\maketitle

\section{Introduction}
Mixed Reality (XR) holds the promise to merge the physical and digital worlds inside a headset, 
allowing users to interact with both real and digital objects in new ways. 
This capability has driven the community to rethink fundamental computing tasks—from retrieving data~\cite{huo2018scenariot,ye2022progesar}, 
to modifying the appearance of physical environments~\cite{lindlbauer2018remixed,marwecki2019mise}, 
or orchestrating embodied interactions~\cite{poupyrev1996go,feuchtner2018ownershift}. 
These advancements, powered by computer vision (CV) and the Internet of Things (IoT), greatly expand the interaction space beyond traditional desktop environments~\cite{anthes2016state}.

Yet a core challenge remains in \emph{embodied interaction}: 
many real-world objects lie beyond arm's reach or are arranged in cramped or occluded spaces, making them difficult to select and interact with.
Imagine trying to select a book on a shelf to retrieve its data (\autoref{fig:teaser}) .
Commercial XR devices often rely on raycasting methods for object selection. 
For example, Quest uses a hand ray + pinch, and the Apple Vision Pro (AVP) requires users to align their gaze with a book and pinch to confirm the selection~\cite{pfeuffer2024design}. 
This process is often cumbersome and error-prone due to the target's small size in view, the natural instability of human gaze~\cite{westheimer1954mechanism}, and hands' tremors and Heisenberg effect \cite{wolf2020understanding}.

Prior techniques like Go-Go~\cite{poupyrev1996go}, Snapmoves~\cite{cohn2020snapmove}, Erg-O~\cite{montano2017erg}, and Ownershift~\cite{feuchtner2018ownershift} attempt to 
extend user reach through non-linear mappings to ease the interaction with remote objects.
However, they often miss-align with the user's mental model and perception, 
breaking the sense of immersion~\cite{gonzalez2024guidelines,gonzalez2017model} and embodiment~\cite{padrao2016violating}. 
Moreover, existing methods struggle with more complex tasks like multi-object selection or switching between group and individual objects~\cite{argelaguet2013survey,mendes2019survey}, often forcing users into laborious single-object targeting or bulky spatial menus~\cite{argelaguet2013survey,wentzel2024comparison}. 
The difficulty of multi-selection with gaze or ray remains an active area of research~\cite{kim2025pinchcatcher}.



Rather than attributing these challenges solely to the interaction techniques themselves,  
we argue that they fundamentally arise from the \textbf{\emph{requirement to interact directly with real-world objects}}, each bound by immutable physical constraints such as size, position, and arrangement.
This may explain why existing interaction techniques perform well in Virtual Reality (VR), 
where objects are fully digital and can be freely manipulated for easier selection.
For instance, users can bring objects closer, enlarge them, or rearrange them entirely, as seen in systems like Poros~\cite{pohl2021poros}, Remixed Reality~\cite{lindlbauer2018remixed}, and Mise Unseen~\cite{marwecki2019mise}.


Our goal is thus to facilitate the interaction with real objects beyond reach in MR 
while preserving the natural mental model of direct manipulation. 
We propose to \emph{\textbf{seamlessly shift the interaction target from the object to its abstract representation}}, or proxy, during selection.
Selecting a proxy is functionally equivalent to selecting the actual object. 
These proxies decouple interaction from physical constraints,
allowing them to be placed in convenient locations 
so users can easily select distant objects and perform advanced selections using familiar direct manipulation gestures. 
This change has zero adaptation cost for users who can continue using their familiar interaction vocabularies.

We embody this concept in \CoolName{}, a prototype that extends the Gaze+Pinch interaction~\cite{pfeuffer2017gaze} on AVP.
When the user pinches, \CoolName{} creates proxies for real-world objects within the user's gaze range, 
using CV and LLMs or predefined digital twins, and places them near the user's hand for easy selection. 
Moreover, enabled by the AI models, \CoolName{} enrich these proxies with the spatial hierarchy and semantic attributes of the real objects.
These structures enable a variety of interactions that were previously difficult or impossible in MR, such as selecting multiple objects by position or attributes, grouping objects, and switching between group or individual selections via zooming--all using familiar gestures.

We demonstrate \CoolName{}'s versatility and effectiveness through several application scenarios, 
including information retrieval in everyday scenarios, building navigation, and drones control. 
An expert review involving 10 XR specialists further underscores \CoolName{}'s utility, usability, applicability, and potential to enhance user experience as a foundational UI design in future MR.
In summary, our main contributions include: 
\begin{itemize}[leftmargin=*]
    \item \textbf{A novel concept and design} for  seamlessly shifting interaction targets from physical objects to their abstract representations.
    \item \textbf{An exploration of advanced interaction features} enabled by our design for interacting with real-world objects in MR.
    \item \textbf{\CoolName{}}, a proof-of-concept \add{open-source\footnote{The project's source code is available at \url{https://github.com/demoPlz/RealityProxy}.}} implementation of our concept extending Gaze+Pinch interactions.
    \item A set of \textbf{application scenarios} that demonstrate the usage of \CoolName{}, supported by \textbf{an expert evaluation} offering insights into its utility, usability, and applicability.
\end{itemize} 
\section{Related Work}
This work focuses on interacting with real-world objects in MR for digital tasks. 
For clarity, we refer MR as a superset of augmented reality (AR) that integrates both physical and digital elements, 
while \emph{XR} encompasses the full spectrum of immersive technologies (VR, AR, MR). 
We review related work in three areas: 
interactions with real-world objects in MR, 
interactions with digital content in XR, 
and abstract representations for interactions.


\subsection{Interacting with Real-World Objects in MR}
A unique aspect of MR is its seamless blending of physical and digital environments. 
Researchers have accordingly explored leveraging the affordances of physical objects to facilitate interactions with digital content, enabling tangible interfaces that harness familiar haptic qualities, geometry, and weight of everyday items for more seamless interaction~\cite{gupta2020replicate, lee2007interaction}. 
For example, \emph{opportunistic interfaces}~\cite{du2022opportunistic, monteiro2023teachable, hettiarachchi2016annexing} enable the re-purposing of everyday physical objects as tangible body for digital content in MR. 
Similarly, PapARVis~\cite{chen2020augmenting, tong2022exploring} exploits the tangibility of paper for interacting with data visualizations, combining direct manipulation on a physical medium with enhanced comfort and intuitiveness. In more dynamic scenarios, actuated robots have been used as reconfigurable tangible embodiments of virtual objects~\cite{le2016zooids}. For example, HapticBots by Suzuki et al.~\cite{suzuki2021hapticbots} employs multiple tabletop-sized shape-changing robots to provide rich haptic feedback in VR.

Another line of work focuses on linking real-world objects 
to digital information to facilitate relevant tasks.
Earlier AR systems often relied on markers~\cite{kan2009applying, ahuja2019lightanchors} or IoT sensors~\cite{ye2022progesar, huo2018scenariot} to identify physical items and retrieve associated digital data. 
Thanks to advances in computer vision (CV), modern systems can now understand scenes and retrieve data using purely visual input~\cite{voulodimos2018deep},
removing the need for explicit markers. 
For instance, in the domain of accessibility, 
researchers have leveraged CV models to retrieve information from the scenes to
enhance various scenarios, including urban commuting~\cite{FroehlichTheASSETS2024}, 
interaction with everyday physical controls~\cite{guo2016vizlens, guo2019statelens, guo2017facade},
cooking~\cite{lee2024cookar}, 
and indoor navigation~\cite{zhao2019designing, chen2025visimark}.
In the context of sports spectatorship, 
CV models have been employed to extract key information from games to enhance the viewing experience~\cite{DBLP:conf/chi/ChenYSLBXP23, chen_augmenting_2022, chen_sporthesia_2022}.
More recently, the advent of Multimodal Large Language Models has further enhanced spatial reasoning and context-awareness in XR. 
For example, XR-Objects~\cite{dogan2024augmented} allows a user to simply touch a physical object to open a context menu for that object.  
GazePointAR~\cite{lee2024gazepointar} leverages eye-gaze and voice input to help users ask questions about objects, and RealitySummary~\cite{gunturu2024realitysummary} provides on-demand augmented summaries of physical documents. 

Building on this growing body of research on interacting with real-world objects for digital tasks, our work focuses on the fundamental interaction aspect in MR. 
Most existing systems rely on conventional interaction methods such as direct bare-hand manipulation or raycasting, which are effective when target objects are within reach and isolated. However, they often fail in scenarios where objects are distant, crowded, or partially occluded. 
This limitation motivates 
us to explore new interaction paradigms
and propose a proxy-based method that supports a broader range of scenarios for interacting with real-world objects in MR.

\subsection{Interacting with Digital Content in XR}
XR interaction has been a rich field of study for decades. 
It has evolved from early controllers-based approaches~\cite{anthes2016state} 
to full hand tracking~\cite{xiao2018mrtouch} and, more recently, 
to techniques combining multiple input modalities such as speech~\cite{bovo2024embardiment}, eye gaze~\cite{pfeuffer2017gaze}, or full-body tracking~\cite{gonzalez2020movebox}. 
In the following, we provide an overview of the major interaction techniques for digital content and discuss some common trade-offs and their relation to our work.
Comprehensive reviews of XR interaction can be found in~\cite{argelaguet2013survey, besanccon2021state, mendes2019survey, wentzel2024comparison, laviola20173d}.

\para{Bare-Hand and Raycasting-Based Methods.}
Early-stage VR interactions primarily fell into two broad categories: bare-hand direct manipulation and raycasting-based methods. 
Bare-hand direct manipulation methods (e.g., \cite{mine1997moving, pei2022hand, hayatpur2019plane,surale2019experimental,ha2014wearhand}) are highly intuitive, leveraging natural motor skills to manipulate virtual content.
However, these techniques are limited by the user's physical arm reach within expansive virtual environments~\cite{gonzalez2024guidelines}.
Raycasting-based methods address this limitation by projecting a virtual ray from the user's hand or controller
(e.g., \cite{baloup2019raycursor, grossman2006design, kopper2011rapid, steinicke2006object, mine1995virtual}). 
While these methods increase the interaction range, they can suffer from inaccuracies due to hand jitter and the Heisenberg effect~\cite{bowman2001using}, 
where the precision of selection becomes uncertain when confirming the selection.
Recognizing the limitations inherent in each approach, 
researchers have explored techniques that combine the best of both worlds. 
These techniques generally fall into the following categories.

\para{Extending Hand Reach.}
One approach focuses on extending the user's hand reach.
A common solution here is remapping technique -- using a non-linear mapping function to increase the user's reach for distant objects while maintaining a linear mapping for nearby interactions (e.g., \cite{cohn2020snapmove, montano2017erg, feuchtner2018ownershift, poupyrev1996go}). 
Another strategy involves separating pointing actions from hand manipulation,
in which the pointing is often performed with the head or eye gaze (e.g., \cite{atienza2016interaction, kyto2018pinpointing, deng2017understanding, clifford2017jedi}). 
For example, the Gaze+Pinch interaction in VR~\cite{pfeuffer2017gaze} uses users' gaze to point and select objects while hand gestures are used to interact and manipulate. 
Our work extends Gaze+Pinch, but moves out of VR and focuses on physical objects in MR. 

\para{Bringing Objects Closer.}
A different approach focuses on bringing virtual objects closer to the user,
either by wrapping the entire space (e.g., \cite{chae2018wall, elmqvist2005balloonprobe, sandor2010egocentric, mine1997moving}) or creating portals to skip the physical distance (e.g., \cite{kotziampasis2003portals, kunert2014photoportals, liu2018increasing, stoev2002application}).
By this, the user can directly manipulate remote objects within arm's reach.
In this context, Poros~\cite{pohl2021poros} introduced by Pohl et al., is conceptually closer to our approach, as it creates manipulable portals (also refereed to as a \emph{proxy}) in VR to enable interacting with distant objects. 
However, our method differs in that we focus on real-world objects in MR, which introduces additional physical constraints. Moreover, instead of using portals, we create proxies for objects to preserve spatial coherence.

\para{World in Miniature.}
In \emph{World in Miniature}~\cite{WIM},
users interact with a smaller version of the virtual world on their hand, 
with actions synced to the original environment. 
Miniature has been explored in VR for different purposes,
including navigation~\cite{pausch1995navigation, pierce2004navigation,abtahi2019m},
collaboration~\cite{piumsomboon2019shoulder},
and interacting with objects beyond arm's reach~\cite{bluff2019don, pierce1999voodoo, pohl2021poros}.
While useful,
accurate selection in Miniature can be challenging, 
particularly for crowded or small objects~\cite{mine1997moving, xia2018spacetime}. 
In our approach, we create abstract representations for real-world objects, which do not necessarily maintain a one-to-one mapping of their physical sizes. 
Instead, these representations are optimized for interaction purposes, ensuring sufficient size and spatial arrangement to facilitate efficient and intuitive user interaction.

\subsection{Abstract Representations for Interactions}
To extend and generalize the principles of direct manipulation~\cite{shneiderman1983direct} beyond simple geometric transformations, 
researchers have proposed developing abstract representations as mediators for interaction.
Notably, Instrumental Interaction\cite{beaudouin2000instrumental} introduced an interaction model in which tools (i.e., commands) are reified as manipulable objects within GUIs. 
This concept has been applied in various scenarios, 
transforming static icons into interactive objects~\cite{bederson1996local, DBLP:conf/chi/BierSPFBCBD94, ciolfi2016beyond}.
Similarly, researchers have also proposed reifying abstract attributes as manipulable objects~\cite{han2020textlets, DBLP:conf/chi/ChenX22, xia2016object, xia2017collection}.

Whereas most of this prior work addresses purely digital GUIs, our goal is to bring a similar concept into MR for real-world objects, which are not as malleable as digital ones. 
Our solution is to shift user interactions from physical objects to their digital proxies.
This approach is particularly suited to MR scenarios where users' goals involve downstream digital tasks rather than altering the object's physical properties. 
\section{Design Considerations and Implementation}
\label{sec:pipeline}

To overcome the physical constraints in MR interactions, 
we propose seamlessly shifting the interaction target from the physical object to an abstract, persistent representation that serves as a proxy.
A proxy is a manipulable object, and selecting a proxy is functionally equivalent to selecting the actual object.
This shift frees interactions with real-world objects from physical constraints such as distance or size.

\begin{figure*}[t]
    \centering
    \includegraphics[width=1\textwidth]{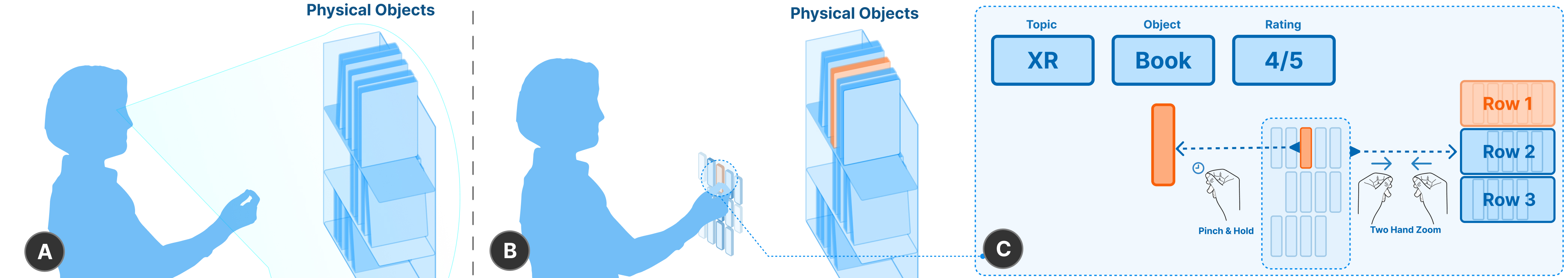}
    \caption{
    Basic workflow of using Reality Proxy to retrieve information of books:
    A) The user gazes at a book and performs a pinch.
    B) The system generates proxies for nearby books and places them within reach.
    C) These proxies enable further interactions, such as filtering by attributes and grouping, using familiar gestures like pinch-and-hold and two-hand zoom.
}
    \label{fig:basic_workflow}
\end{figure*}

We embody this concept in \CoolName{},
a proof-of-concept system that extends the Gaze+Pinch interaction~\cite{pfeuffer2017gaze} 
to showcase how our approach can enhance SOTA MR interactions. 
\autoref{fig:basic_workflow} depicts a basic use case of \CoolName{} for retrieving information about a book on a shelf. 
The user gazes at the book and performs a pinch gesture (\autoref{fig:basic_workflow}a).
\CoolName{} creates proxies for the nearby books, 
places them within reach,
and provides visual feedback on the books as the user interacts with the proxies (\autoref{fig:basic_workflow}b).
Beyond single selections, 
these proxies enable further
interactions, such as filtering by attributes and grouping, using familiar gestures like pinch-and-hold and two-hand zoom (\autoref{fig:basic_workflow}c). 
More details on these interactions can be found in Sec.~\ref{sec:feats}.

The process of \CoolName{} consists of three main interaction steps:
1) \textbf{activating}, 2) \textbf{generating}, and 3) \textbf{interacting} with the proxies.
Next, we delineate the design considerations, our current implementation, and potential future improvements for each step. 

  \begin{figure*}[b]
    \centering
    \includegraphics[width=1\textwidth]{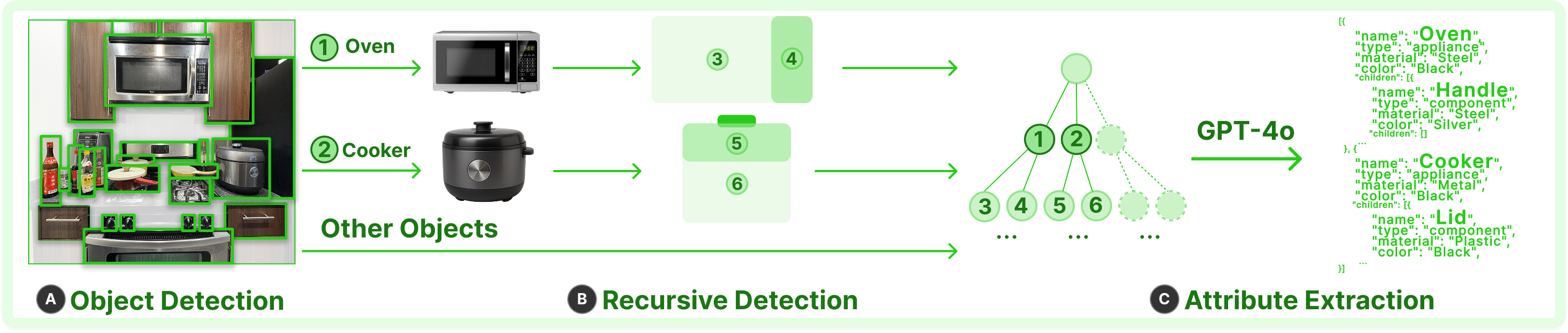}
    \caption{An AI-driven pipeline for extracting the spatial hierarchy and semantic attributes of objects in the scene. The pipeline recursively detects objects and organizes them into a tree structure based on spatial containment. For each object, GPT-4o generates descriptive attributes. The resulting JSON captures the AI's structured understanding of the scene.
    }
    \label{fig:pipelineA}
  \end{figure*}

\subsection{Activating Proxies: Capturing Hierarchical and Semantic Scene Structure}
\label{sec:activating}

\CoolName{} is designed to be activated as part of the existing Gaze+Pinch interaction without adding extra steps:
when the user pinches to confirm selection,
it automatically detects the real-world objects within the user's gaze range 
and abstracts them into interactive proxies at the user's hand.
This method introduces no disruption: 
if the user successfully selects the target object (the default one targeted by the user's gaze), 
they can proceed with their intended actions. 
Otherwise, they can refine the selection using the proxy at hand, as described in Sec.~\ref{sec:feats}.

To support generating proxies for real-world objects, 
\CoolName{} extracts structural information from the scene during activating. 
We develop an AI-driven pipeline (\autoref{fig:pipelineA}) to achieve this. 
Specifically, we extract both the spatial hierarchy and semantic attributes of objects as follows:

\para{Detecting Objects in Hierarchical Structures.}
A critical challenge here is enabling users to interact with targets at varying levels of granularity. 
For instance, in~\autoref{fig:basic_workflow}a, 
a user might want to select an entire bookshelf, 
or conversely, only a single book. 
To accommodate these varied needs,
we employ a hierarchical detection strategy powered by DINO-X~\cite{ren2024dinoxunifiedvisionmodel}, which excels at open-vocabulary object detection and segmentation. 
The process includes three major steps:
\begin{enumerate}[leftmargin=*]
    \item \emph{Gaze Region Extension}: The user's gaze range is slightly expanded using a predefined spatial threshold (e.g., 20 cm) to capture additional contextual information.

    \item \emph{Object Detection}: 
    The video frame of the user's extended gaze region is fed into the model to produce 2D bounding boxes (bboxes), pixel masks, and semantic labels for all detected objects (\autoref{fig:pipelineA}a). 
    We sort the bboxes in descending order of size and remove duplicates whose Intersection over Union (IoU) exceeds 0.75. The remaining boxes constitute \emph{Level 1} objects.
     
    \item \emph{Recursive Detection}: 
    Each Level 1 object is cropped from the image and reprocessed by the model to detect finer subcomponents, resulting in \emph{Level 2} objects (\autoref{fig:pipelineA}b). This step is repeated recursively until no further objects are detected or the bounding box size falls below a specified threshold.
    
\end{enumerate}
The final output is a naturally hierarchical structure, aligned with human conceptual understanding of scenes~\cite{armeni20193d}. 
Note that during interactions, only one level of objects will be shown by default.

\para{Extracting Semantic Attributes of Each Object.}
To further support downstream interactions, we extract descriptive semantic attributes for each detected object.
These attributes allow users to filter or group proxies by attributes,
even when the objects are not physically co-located.
For example, one might group books by topics or colors.
Such attribute-based interactions are common in traditional desktop interfaces~\cite{xia2016object, xia2017collection}.

We implement this by sending cropped images of detected objects from the previous step
to GPT-4o for attribute extraction.
This results in a hierarchical and semantically rich scene representation  (\autoref{fig:pipelineA}c) in a JSON format,
where each node encodes both containment relationships and semantic attributes.
This structure forms the basis for the interactions supported by \CoolName{}.

\para{Future Improvements: Toward More Advanced AI or Beyond Vision-Only Solutions.}
In our current implementation, object detection is performed continuously before the user executes a pinch gesture, while semantic attribute extraction is performed asynchronously after the pinch. 
However, our purely vision-based pipeline cannot handle objects that are fully occluded.
Moreover, the quality of the scene representation directly determines the user's interaction space—it should capture all relevant objects while excluding irrelevant ones.
We further discuss potential solutions to these critical challenges in Sec.~\ref{sec:discussion}.
Nevertheless, if digital twins or precise location data from external sensors are available, our system can generate proxies for otherwise unrecognized objects, as exemplified in Sec.~\ref{sec:building} and Sec.~\ref{sec:drone}.

\subsection{Generating Proxies: Preserving Spatial Relationships}
\label{sec:generating}

In this step, we translate the hierarchical, semantic representation of the scene from the previous step into \emph{proxies}—user-manipulable objects. 
By default, we generate proxies only for Level 1 objects within the user's extended gaze region.
These proxies preserve their relative spatial relationships to one another.
Each proxy is manipulable via standard gestures—such as long-press and two-hand zoom—and remains persistent in place even after the user releases the pinch. 
This stable arrangement allows users to easily refine their selections by directly manipulating nearby proxies, rather than realigning their gaze. It also supports more advanced interactions introduced in Sec.~\ref{sec:feats}.

\begin{figure}[h]
\centering
\includegraphics[width=1\columnwidth]{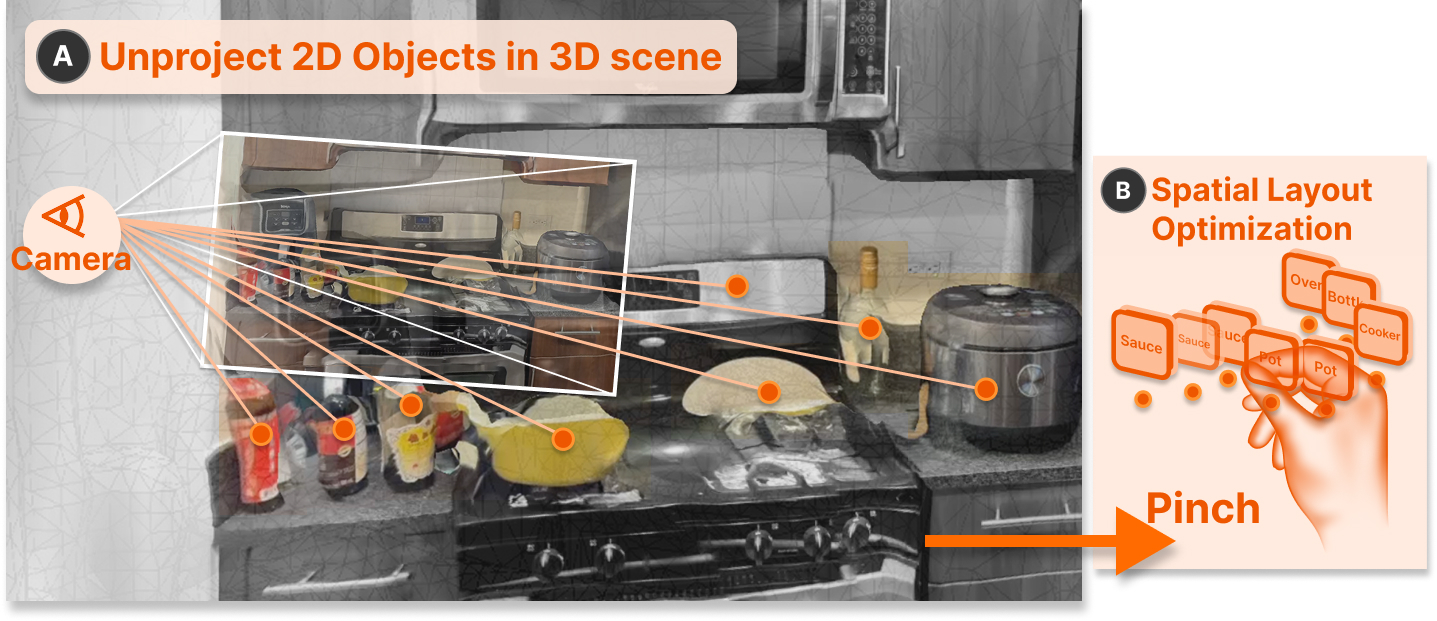}
\caption{A) Obtaining the 3D positions of detected objects via raycasting.
B) Generating a proxy layout that preserves their relative spatial relationships.
}
\vspace{-2mm}
\label{fig:pipelineB}
\end{figure}

Since proxies serve as abstract representations for interaction purposes only, their physical size is not critical. Therefore, we represent each proxy as a fixed-size, rectangular 3D object (\autoref{fig:pipelineB}b) in our implementation.
The key design requirement, however, is to preserve their relative spatial relationships, enabling users to interact with the scene in a spatially coherent manner through gestures.
We achieve this through the following two steps.

\para{Obtaining 3D Positions via Raycasting to the Scene Mesh.} 
While 3D object detection~\cite{yin2021center} could be used, it would introduce additional computational overhead and uncertainty. Instead, we reuse the object detection results from the activation phase and perform a raycast from the user's head to the center of each 2D bbox (\autoref{fig:pipelineB}a). 
We then approximate the object's 3D position
using the position where each ray intersects the scene mesh (provided by the headset's operating system).
This approach offers a lightweight and efficient alternative for obtaining approximate 3D locations.

\para{Preserving Relative Spatial Relationships through Constraint-Based Layout Optimization.} 
Directly placing proxies at the raw 3D positions—or simply scaling those positions—can result in impractically large layouts or excessive gaps between proxies. 
Instead, we rescale and reposition objects to reduce their absolute distance 
while preserving their relative spatial layout. 
To achieve this, we apply a constraint-based layout optimization method.
Specifically, we generate spatial constraints (e.g., ``Object A is to the left of Object B'' $\rightarrow x_{A} < x_{B}$) from the objects' 3D positions, 
then minimize deviations in inter-object distances, 
ensuring at least 0.5 cm spacing between objects in our prototype. This process is solved using Z3~\cite{de2008z3}.
We acknowledge that advanced techniques such as qualitative spatial reasoning~\cite{freksa1991qualitative} could also be applied here, and we plan to explore them in future work.

\para{Future Improvement: Handling Complex Object Shapes and Layouts.} 
For simplicity, our current implementation assumes that each object is convex and does not intersect with others, allowing for a fixed-size 3D representation. 
However, real-world objects may be non-rigid or highly irregular, requiring more sophisticated and context-aware strategies, as discussed in Sec.~\ref{sec:discussion}.
Our current layout generation method is intuitive for typical scenes but may not handle edge cases well. Moreover, the static layout may not accommodate individual user preferences and, as observed in our user study, can sometimes lead to inaccuracies during interaction.
Future work could explore these advanced layout generation techniques 
and integrate other disambiguation methods~\cite{argelaguet2013survey} to improve the accuracy, robustness, and personalization of the system.
\add{To enhance usability, future versions could also attach a thumbnail of each physical object (reused from segmentation) onto its proxy, which would help disambiguate proxies even when their corresponding real objects are out of view.
}

  \begin{figure}[h]
    \centering
    \includegraphics[width=1\columnwidth]{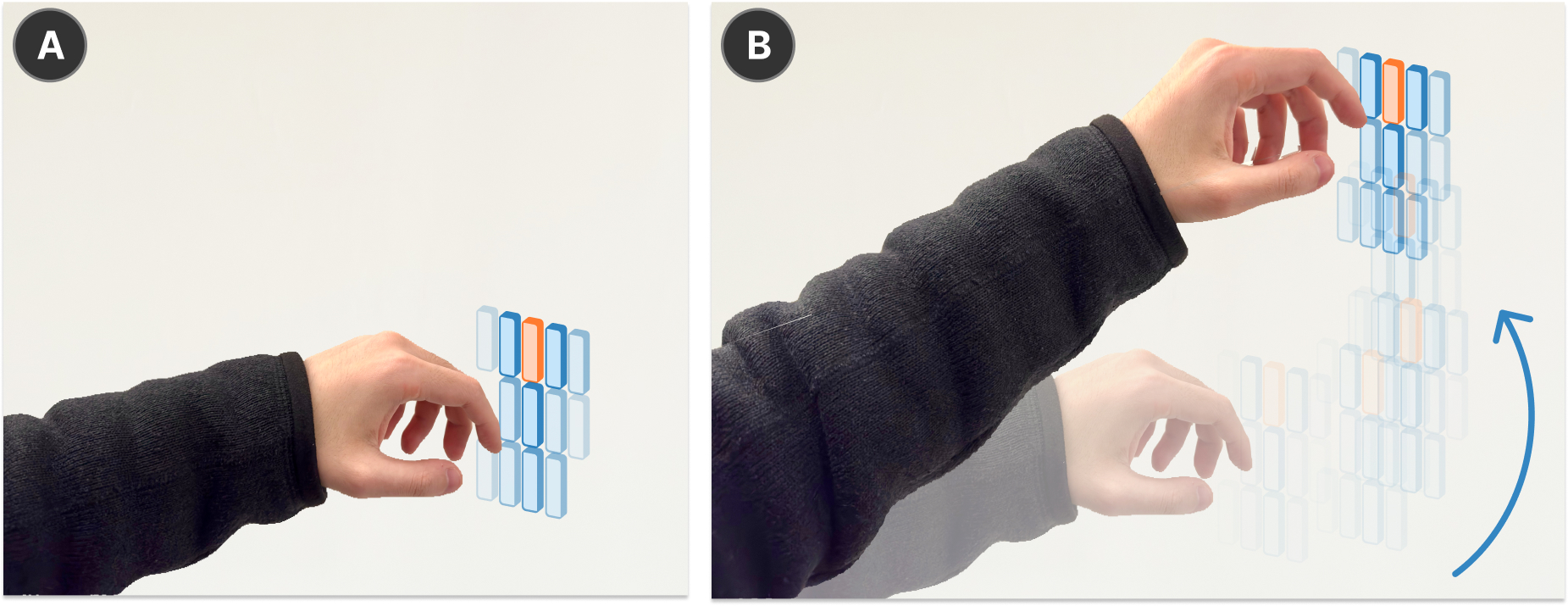}
    \caption{A lazy-follow mechanism that positions proxies near the user's hand
    for easy access without looking down.
    }
    \label{fig:pipelineC}
  \end{figure}

\subsection{Interacting With Proxies: Maintaining Real-World Focus}
\label{sec:interacting_proxies}
To allow the user to primarily focuses on the physical objects,
\CoolName{} displays key visual feedback directly on the physical objects when interacting with a proxy. 
For instance, when an object is selected, it is highlighted in a bright color, and the corresponding proxy is also highlighted, providing dual feedback. 

To ensure proxies remain easily accessible without requiring constant visual attention,
we apply a lazy-follow~\cite{unityLazyFollow} mechanism that positions proxies near the user's hand (\autoref{fig:pipelineC}).
When the hand remains within a certain threshold, the proxy stays stationary.
If the hand moves beyond this range, the proxy smoothly follows—keeping it within reach without reacting to minor hand jitters.
This design reduces the need for the user to look down to locate the proxy and supports fluid transitions between focusing on the real world and briefly glancing at the proxy.
Together, these mechanisms allow proxies to function like remote controls—extending interaction capabilities without disrupting spatial attention.

\para{Future Improvements: Decoupling Interaction From Physical Collision.}
Currently, interactions with proxies are triggered through physical collision—when the user's finger intersects with a proxy.
However, when users focus on physical objects, minor unintended hand movements can lead to inaccurate interactions.
To address this, a more robust approach would decouple interaction from strict physical collision.
Once the user initiates a gesture (e.g., a pinch), the system could use a machine learning model to infer the intended target based on the hand's relative position and trajectory, rather than requiring direct proxy contact.
This approach could enable more accurate interactions, especially when users operate by ``feel'' while visually attending to the physical world.

\section{Fluid Interactions Enabled By \CoolName{}}
\label{sec:feats}
\setlength{\intextsep}{0pt}%
\setlength{\columnsep}{10pt}%

By shifting the interaction target from physical objects to its proxies, 
\CoolName{} not only mitigates the physical constraints inherent in direct object interaction
but also enables novel MR interactions that were previously challenging or impossible. 
Below, we exemplify these possibilities through selected advanced interactions, 
such as multiple selections, semantic and hierarchical grouping, and leverage physical affordances.
Our goal is to demonstrate how the proxy-based approach fundamentally expands the interaction design space for real-world objects in MR.

\newpage
\subsection{Skim and Preview Objects}

\begin{wrapfigure}{r}{0.43\columnwidth}
  \centering
    \includegraphics[width=0.43\columnwidth]{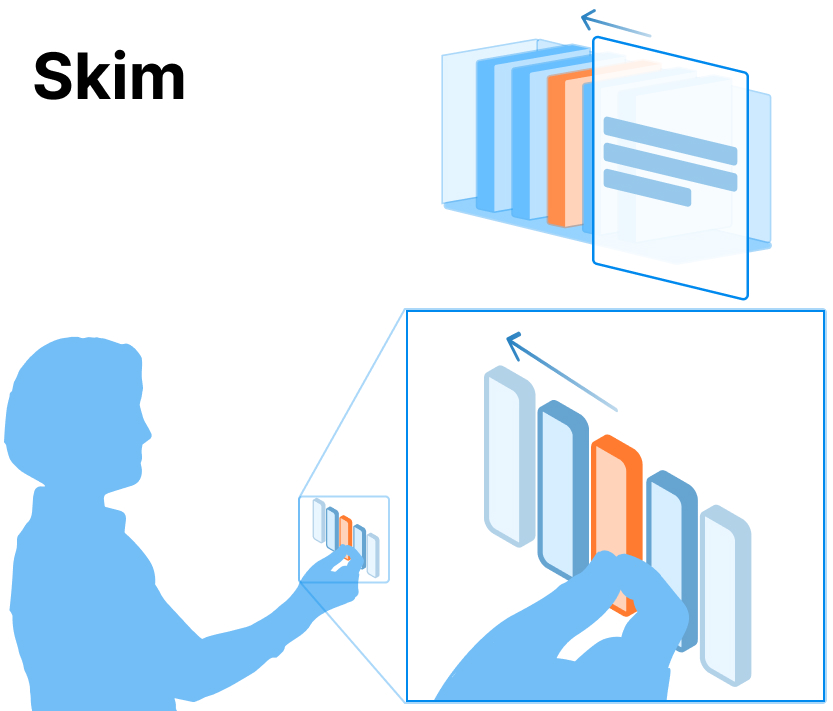}
\end{wrapfigure}

Users can efficiently skim and preview object information by keeping their finger tapped while sliding across multiple proxies. 
For instance, 
users can rapidly browse book information by sliding their finger across proxies.
\CoolName{} will display the book's information using an attribute panel near the book.
This was previously difficult with gaze-based or raycasting interactions,
as pop-up preview windows can unintentionally attract the user's gaze~\cite{flanagan2003action},
and precise ray alignment with distant objects is often unstable to control.

\subsection{Multiple Selection through Brushing}

\begin{wrapfigure}{r}{0.43\columnwidth}
  \centering
    \includegraphics[width=0.43\columnwidth]{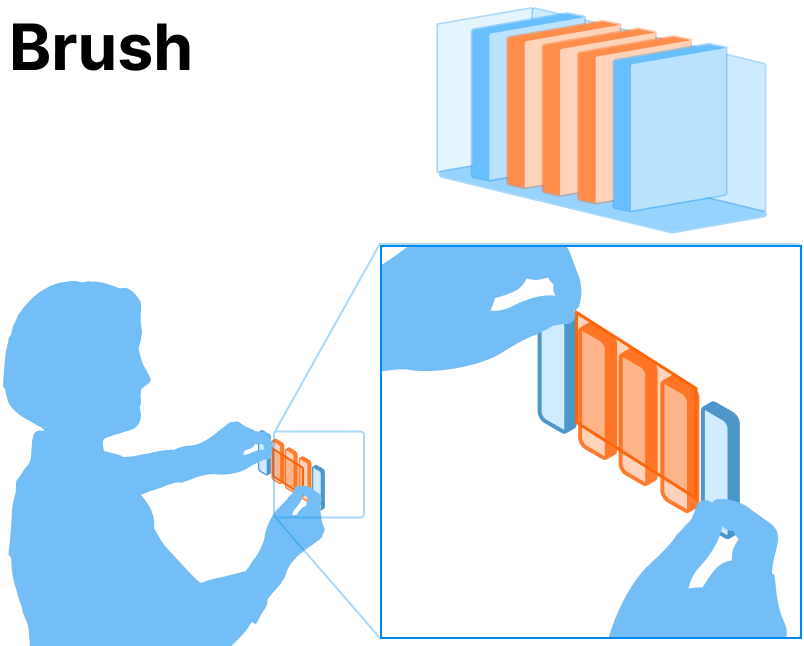}
\end{wrapfigure}

Selecting multiple objects with traditional MR interactions is often challenging, especially when objects are distant, as users typically need to select them one by one using ray-casting methods. 
\CoolName{} simplifies this task by bringing targets closer to the user's comfortable interaction zone.
The user can brush over multiple proxies
by initiating a select region, with a two-hand pinch gesture,
and then expand its size and position by moving their hands apart. 
The corresponding objects are selected and highlighted in place, 
allowing users to remain visually focused on the  objects while brushing.

\subsection{Filtering Objects by Attribute}

\begin{wrapfigure}{r}{0.43\columnwidth}
  \centering
    \includegraphics[width=0.43\columnwidth]{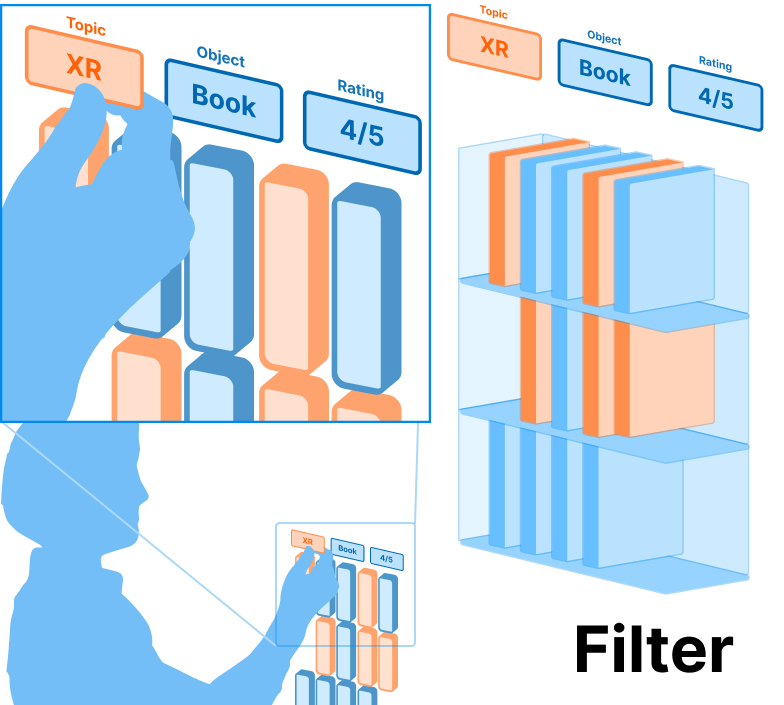}
\end{wrapfigure}

Users frequently need to select subsets of items from larger sets.
In traditional desktop interfaces, widgets such as dropdown lists often facilitate this.
In MR, however, using such widgets poses challenges: placing them near distant objects makes them difficult to interact with due to small size, whereas placing them closer requires users to repeatedly shift visual attention between widgets and physical objects.

With \CoolName{}, digital widgets themselves can also be represented as proxies placed conveniently near the user's hand, creating a unified and seamless MR interaction experience for both physical objects and digital UI elements. 
Building on this capability, we introduce an attribute-filtering mechanism, allowing users to efficiently filter objects based on their semantic attributes extracted in Sec.~\ref{sec:activating}.

Specifically, users can pinch and hold an object's proxy to pin its attribute panels, 
activating a secondary selection mode. 
\CoolName{} then creates proxies for these attribute panels, 
positioning them near the object's proxy.
The user can slide their finger over to an attribute proxy, similar to skimming proxies,
to select all other objects sharing the same attribute value.

\newpage
\subsection{Interactions Leveraging Physical Affordance}
\begin{wrapfigure}[16]{r}{0.45\columnwidth}
  \centering
    \includegraphics[width=0.45\columnwidth]{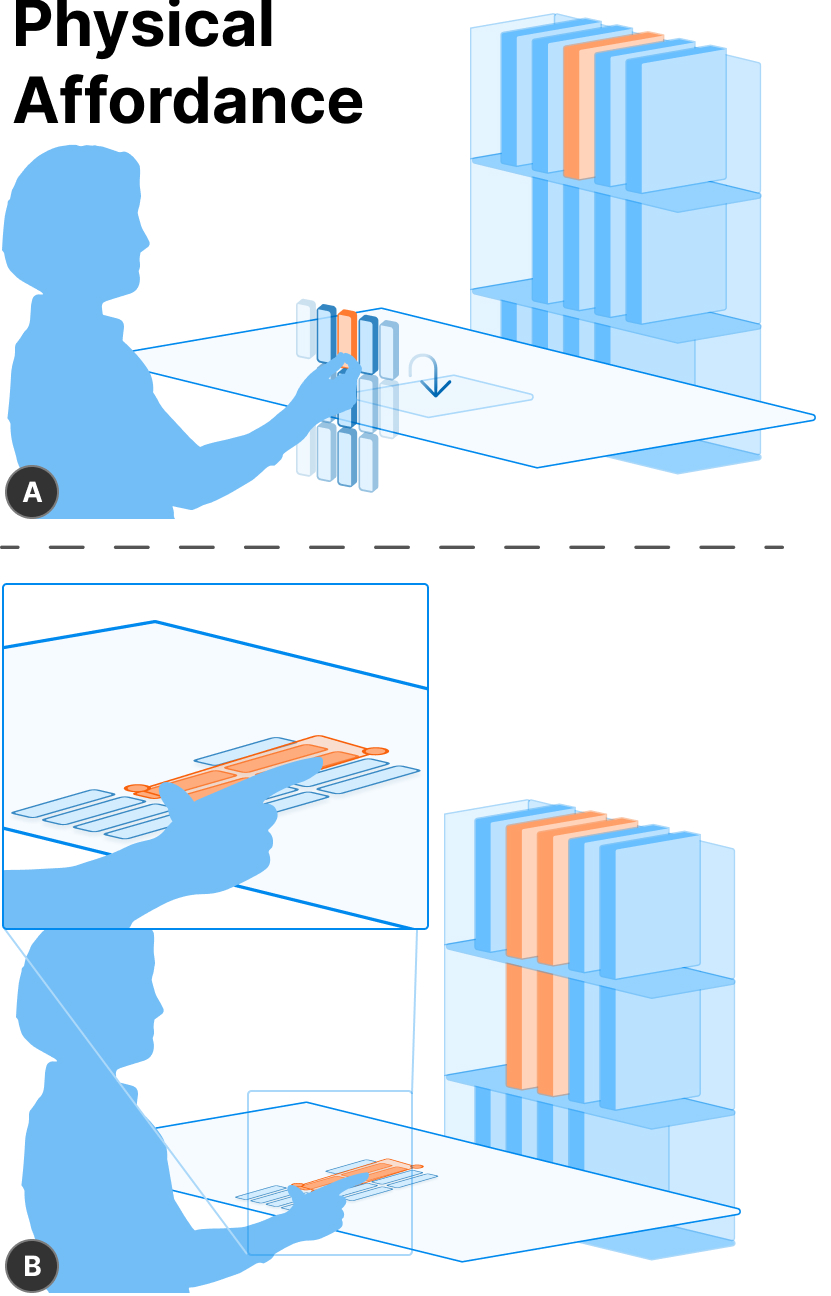}
\end{wrapfigure}

Proxies offer a unique opportunity to leverage real-world physical affordances
to facilitate intuitive interactions. 
For example, 
proxies placed onto physical surfaces (e.g., tables) can transform these surfaces into natural touchpads. 
Users can thus interact with real-world objects using familiar gestures of touch-based devices, 
such as dragging fingers across the surface to select multiple objects, 
spreading fingers apart to expand selection, or retracing their paths to adjust their selections.


Beyond finger-based gestures,
a pen or stylus can be used to enable pen-based interactions.
For example,
users can lasso-select multiple objects by drawing around the proxies or sketch directly onto the surface to interact with objects.
This combination of digital proxies and tangible surfaces bridges MR interaction techniques with familiar touchscreen modalities, reducing interaction gap between different devices.


\subsection{Grouping Objects by Semantic Attributes}

\begin{wrapfigure}{r}{0.46\columnwidth}
  \centering
    \includegraphics[width=0.46\columnwidth]{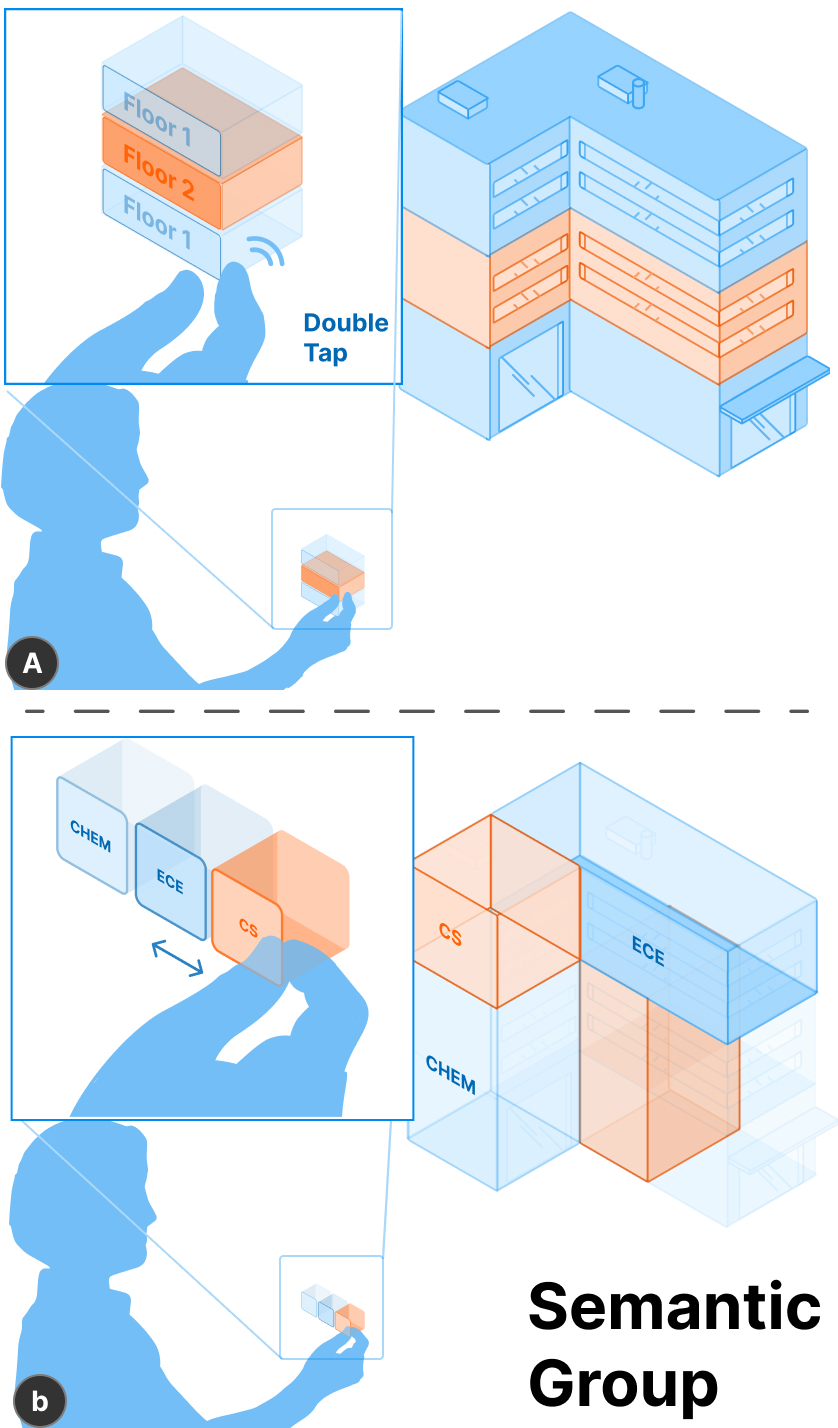}
\end{wrapfigure}

Beyond spatial containment relationships, 
humans regularly organize and group objects by semantic attributes, such as type, function, or topic—reflecting internal cognitive and perceptual categorization processes. 
Leveraging the semantic attributes extracted in Sec.~\ref{sec:activating}, \CoolName{} supports intuitive semantic grouping of objects by shared attribute values.
Specifically, users can trigger semantic grouping by double-tapping a proxy.
Objects sharing the same attribute are highlighted with the same color. 
Simultaneously, 
nearby proxies are dynamically replaced by proxies representing other objects within the same attribute group. 
This dynamic re-representation simplifies identifying and interacting with semantically related objects. 
For example, as illustrated in the right Figure, double-tapping a room grouped by floor reconfigures the view to group rooms by department, enabling users to shift between different semantic organizations of the same physical space.
Users can subsequently apply all other available single-object interactions to these newly grouped proxies.

\newpage
\subsection{Grouping Objects via Spatial Zooming}

\begin{wrapfigure}{r}{0.45\columnwidth}
  \centering
    \includegraphics[width=0.45\columnwidth]{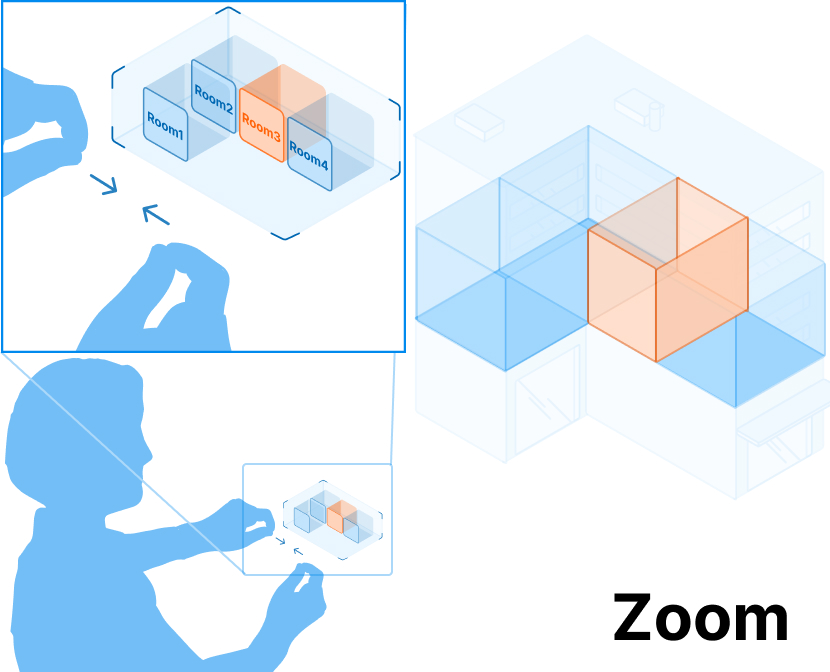}
\end{wrapfigure}

Desktop software (e.g., Adobe Illustrator, PowerPoint)
frequently allows users to group objects hierarchically to facilitate efficient organization and interaction.
Similarly, in real-world scenarios, objects often exist within hierarchical spatial arrangements—books organized into shelves, rooms grouped into floors, or groceries sorted by categories. 
These spatial groupings align closely with human cognitive models of spatial relationships and structures~\cite{Tversky_2008_SpatialCognition}.
Because \CoolName{} captures and maintains these hierarchical spatial structures during object detection, users can intuitively navigate among hierarchical groups using a two-handed zoom gesture. For example, users can fluidly zoom to group rooms within a building into their respective floors, enabling seamless transitions between detailed and overview perspectives aligned with task-specific needs.

\mbox{}
\subsection{Creating Custom Groups for Objects}

\begin{wrapfigure}{r}{0.43\columnwidth}
  \centering
    \includegraphics[width=0.43\columnwidth]{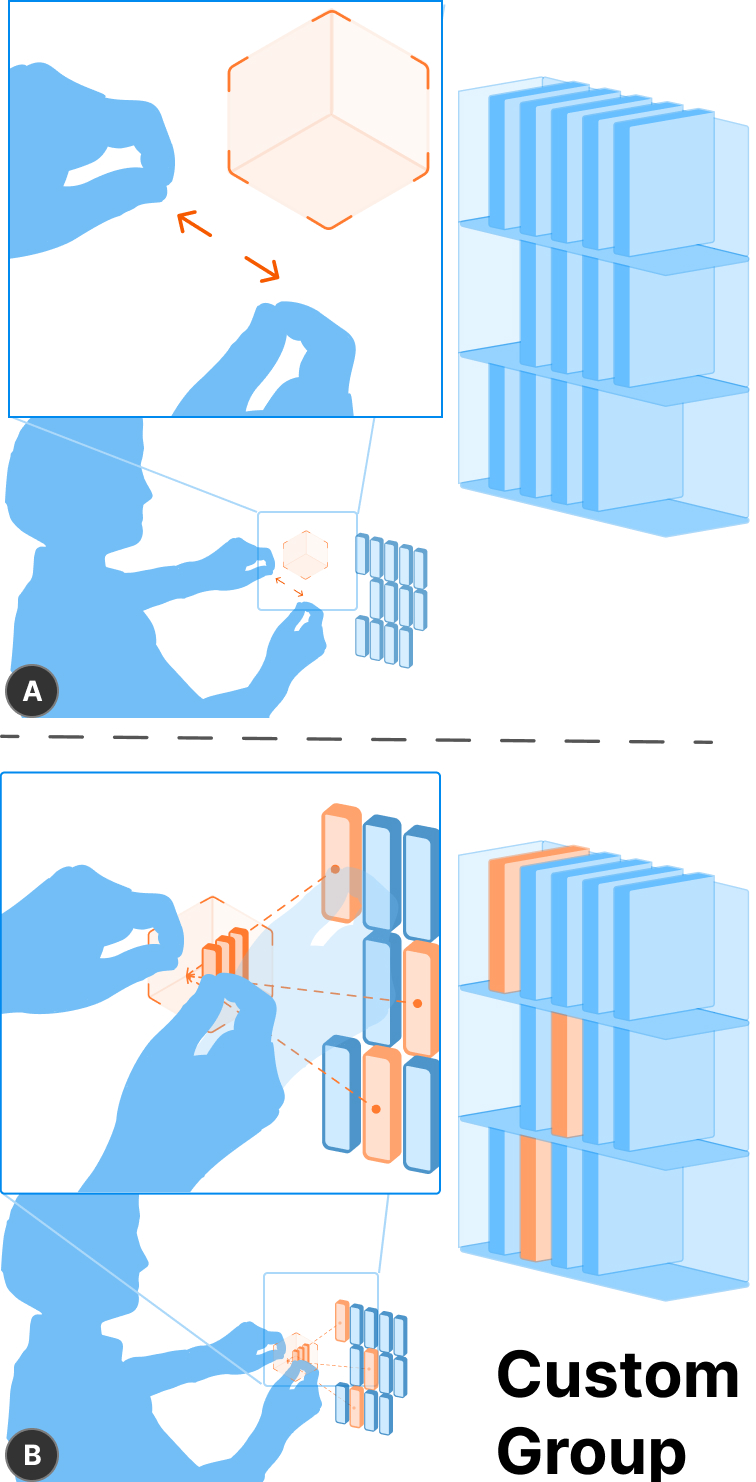}
\end{wrapfigure}

In many scenarios, users need to create custom, task-specific object groups.
\CoolName{} supports custom grouping through intuitive gestures. 
Users can creat a cube-shape group container by performing a brushing gesture—previously introduced for selection—in empty space. 
Upon releasing the gesture, the cube becomes persistent.
To add objects, users pinch and hold the container with one hand and tap proxies with the other, cloning them into the group.
Additionally, users can employ the two-handed zoom gesture to shrink the container into a single proxy, which support all single-proxy interactions.
This flexible grouping feature particularly benefits tasks requiring the collection and manipulation of arbitrary sets of objects-such as summarizing data and calculating totals.
 


\setlength{\intextsep}{\oldintextsep}%
\section{Applications}
To demonstrate the applicability of our method, we developed three proof-of-concept applications: retrieving object information in everyday environments (e.g., office, kitchen), navigating rooms in a building, and controlling multiple drones.

\subsection{Scenario1 - Everyday Information Retrieval}
\label{sec:office}

We developed an everyday information retrieval app based on the pipeline introduced in Sec.\ref{sec:pipeline} using an AVP.
The app allows users to scan objects 
in an office or kitchen and retrieve associated data from tabular datasets. 

\begin{figure}[h]
  \centering
  \includegraphics[width=\columnwidth]{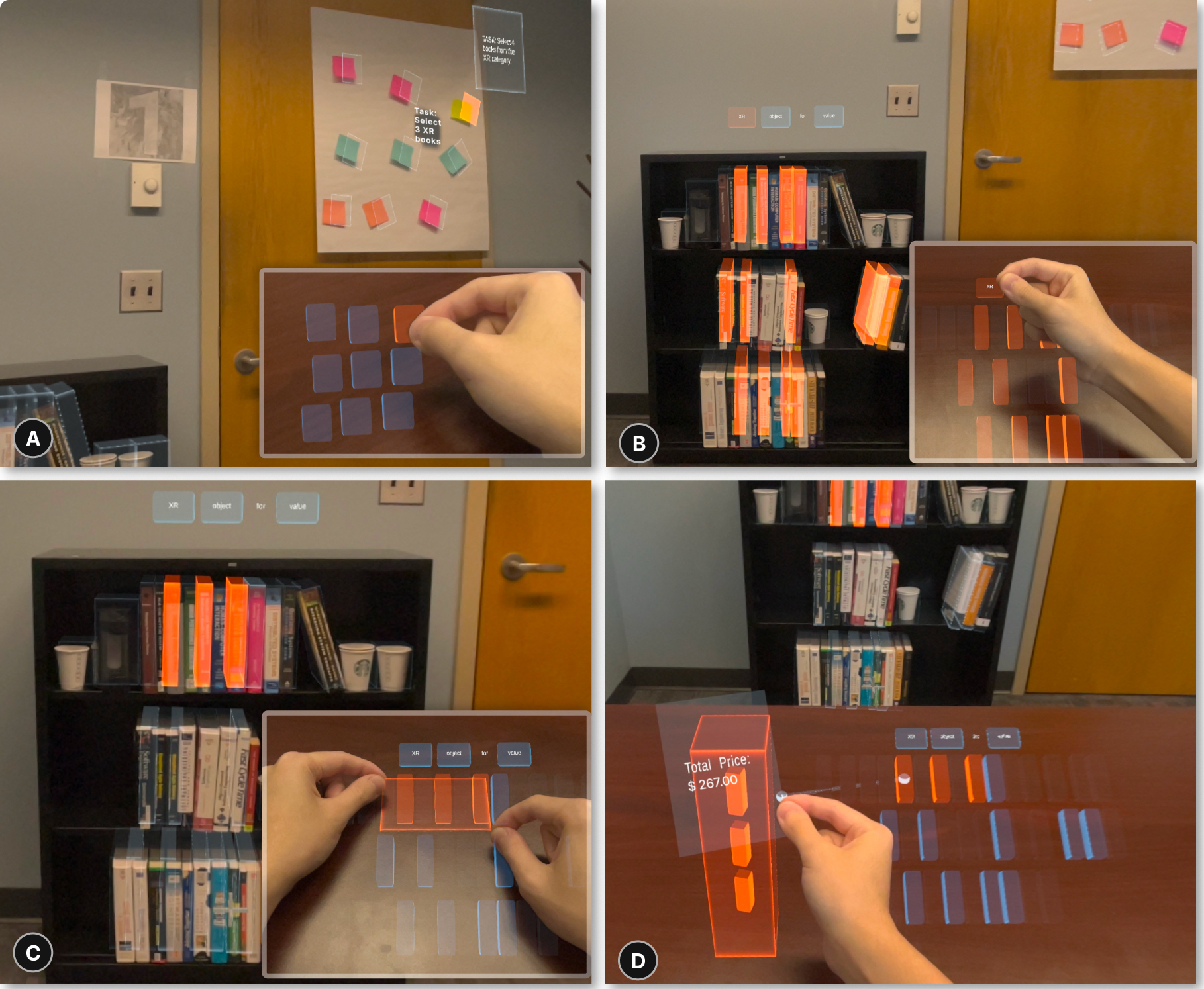}
  \caption{Scenario - retrieving object information in an office.
  }
  \label{fig:app1_book}
\end{figure}

\noindent
\autoref{fig:app1_book} illustrates an example where Bob, a professor, uses the app to share the prices of a few XR books with his student. Bob first looks at the sticky notes on his whiteboard and pinches to initiate selection (\autoref{fig:app1_book}a). 
The app generates proxies for the notes near his hand, enabling him to skim through their contents by moving his hand.
Upon finding the note indicating ``the two XR books on the left on the first row'', 
Bob turns to his bookshelf and pinches again to generate proxies for the books. 
He filters for XR books by long-pressing a book proxy and sliding to select the XR attribute (\autoref{fig:app1_book}b). The system highlights all XR books in view. 
Bob then uses a two-hand brush gesture to select the two target books (\autoref{fig:app1_book}c).
Finally, Bob creates a group by brushing an empty area to form a cube container and adds the selected books into it. The system automatically calculates and displays the total price of the selected books (\autoref{fig:app1_book}d). 

\begin{figure}[h]
  \centering
  \includegraphics[width=\columnwidth]{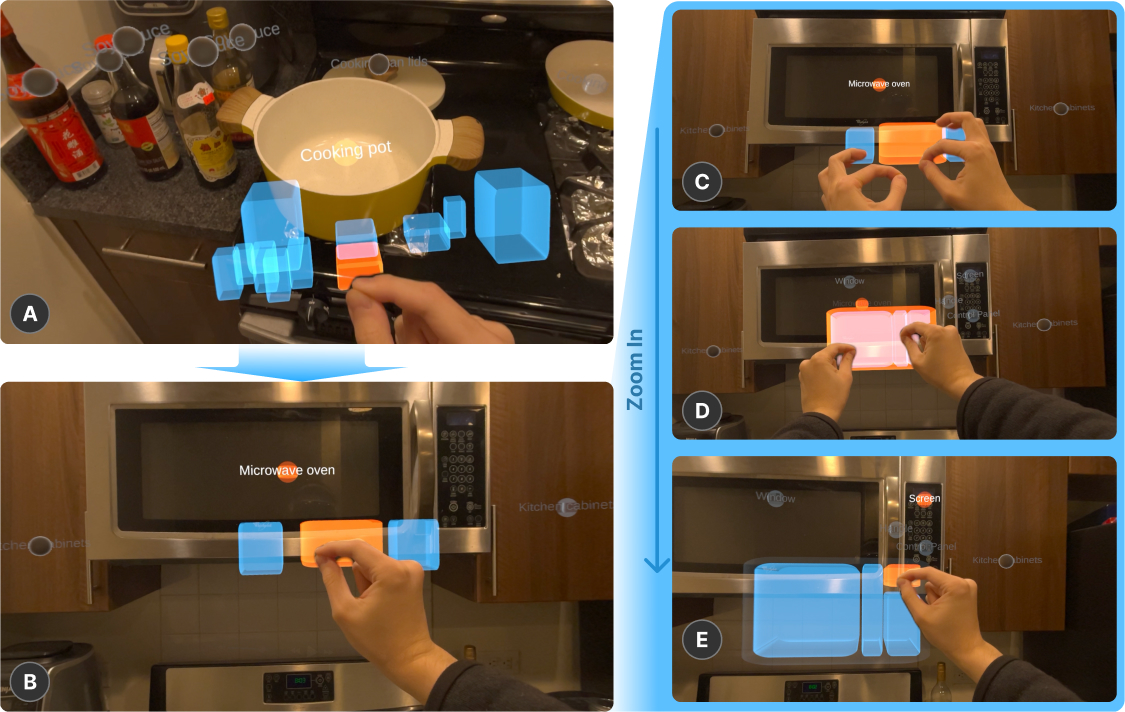}
  \caption{Scenario - interacting with scattered kitchen objects.}
  \label{fig:app1_kitchen}
\end{figure}

\autoref{fig:app1_kitchen}a presents another example of an everyday scenario, 
demonstrating how \CoolName{} handles complex object layouts in a kitchen environment. 
Unlike the office scenario, objects here are placed more randomly. 
\autoref{fig:app1_kitchen}b–d further shows how the user can zoom to interact with objects in different levels of granularity—for example, selecting the entire microwave, its door, or  buttons on its control panel. This demonstrates how \CoolName{} enables users to interact with densely packed objects in real-world scenes.

\subsection{Scenario 2 — Building Navigation}
\label{sec:building}

\noindent\CoolName{} enables interaction with large-scale real-world objects, such as buildings, even when they are invisible or nested within hierarchical structures. 
To demonstrate this, we developed a building navigation app using a predefined digital twin, replacing the AI scene understanding components in the first two steps (i.e., Activating and Generating) of our pipeline.

\begin{figure}[!ht]
  \centering
  \includegraphics[width=\columnwidth]{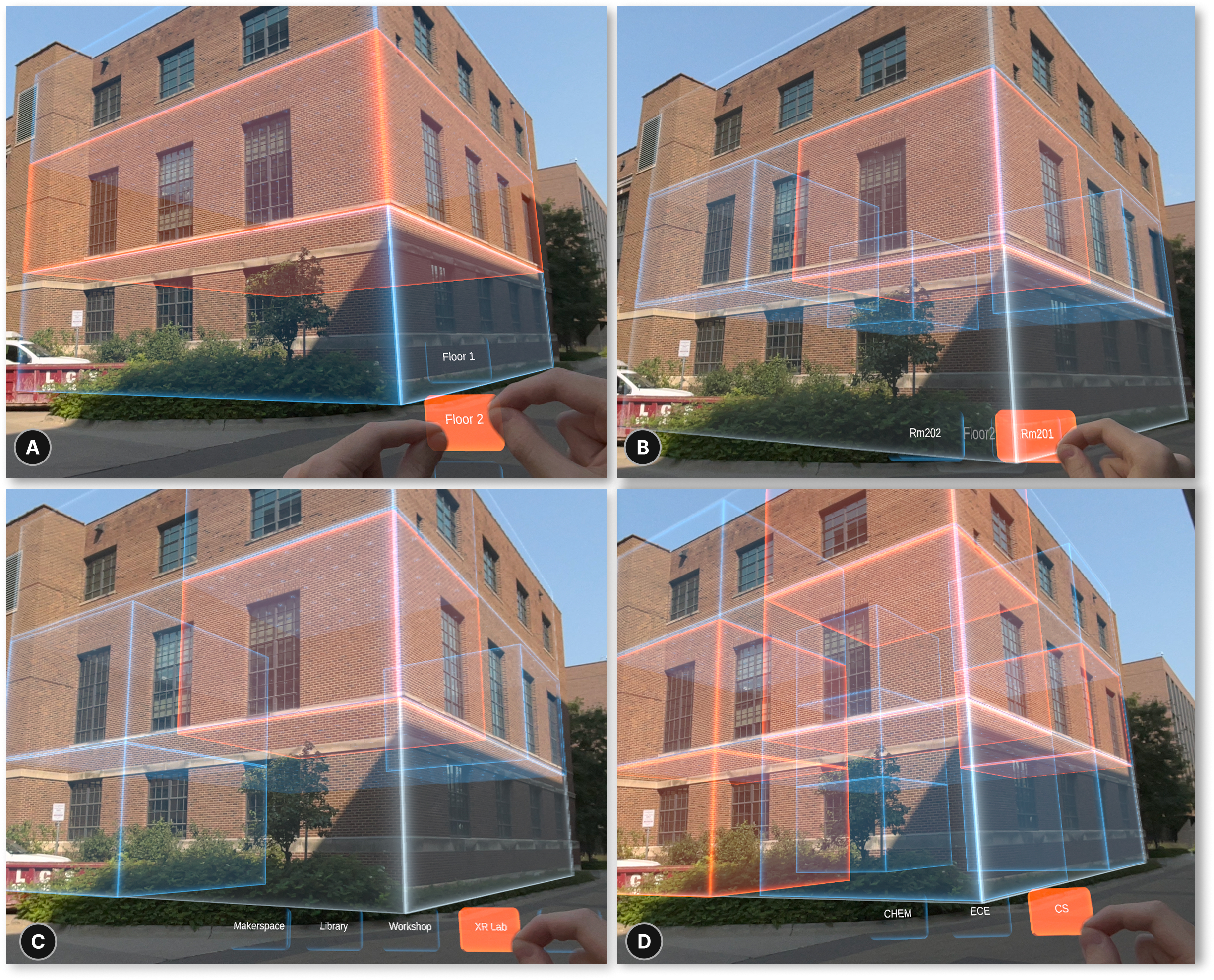}
  \caption{Scenario - Navigating a building using proxies.
  }
  \label{fig:app2_building}
\end{figure}

As shown in \autoref{fig:app2_building}a, the user first selects a floor by pinching on the building. The system reveals the building's structure with an x-ray visual effect. 
The user can then zoom into a specific floor and skim across rooms by sliding their fingers (\autoref{fig:app2_building}b). 
This overcomes the limitations of traditional raycasting methods, which struggle with occlusion and overlapping rooms.
The user can further group rooms by semantic attributes. 
For example, double-tapping a room groups all rooms belonging to the same department and highlights them (\autoref{fig:app2_building}c). 
Zooming out allows the user to view different labs and explore their spatial distribution (\autoref{fig:app2_building}d).

This scenario demonstrates how \CoolName{} supports fluid exploration of nested spatial structures and semantic groupings, enabling efficient navigation and interaction in large scale building layouts — a task that is challenging for traditional MR interactions.

\begin{figure*}[t]
  \centering
  \includegraphics[width=1\textwidth]{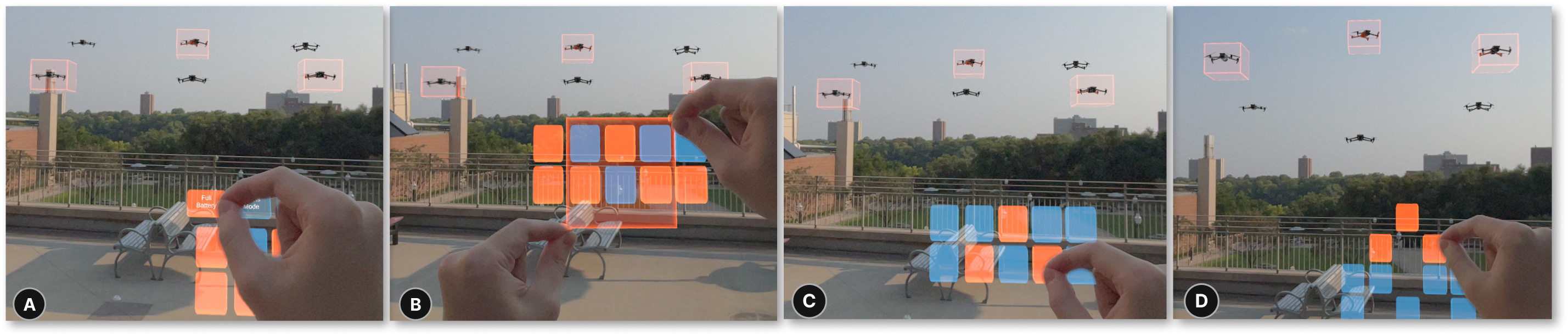}
  \caption{Scenario - Controlling multiple drones using proxies.}
  \label{fig:app3_drone}
\end{figure*}

\subsection{Scenario 3 — Controlling Drones}
\label{sec:drone}

\CoolName{} also allows control of dynamic real-world objects. 
To demonstrate this, we developed an MR-based drone control app, 
which replaces the AI scene understanding component with embedded trackers in the drones.

As shown in \autoref{fig:app3_drone}, the user can select drones and issue commands based on spatial position or semantic attributes. For example, by skimming over the drones, the system displays attribute panels in place, allowing the user to filter for drones with a full battery via a pinch-and-hold gesture (\autoref{fig:app3_drone}a). 
The user can then brush to select drones located on the left (\autoref{fig:app3_drone}b), and command them to move forward using a pinch-and-move gesture (\autoref{fig:app3_drone}c–d).

This scenario demonstrates how \CoolName{} enables direct and efficient multi-drone control in MR, a task that is challenging for traditional techniques such as raycasting on moving objects or minimap-based remote controls~\cite{huo2018scenariot}. 
By supporting selection based on both spatial position and attributes, \CoolName{} simplifies interaction with dynamic objects in physical environments.

\section{Expert Evaluation}

We developed \CoolName{} as a technology probe to explore how proxy-based interactions might mitigate physical constraints in MR. 
We acknowledge that \CoolName{} may, at first place, includes initial usability challenges for users who are unfamiliar with immersive devices like the AVP,
and there is currently no direct comparison baseline. 
Consequently, we conducted an expert evaluation, similar to prior studies~\cite{han2020textlets, DBLP:conf/chi/ChenX22, xia2016object, xia2017collection}, focusing on qualitative insights regarding \CoolName{}'s usefulness, limitations, and potential applications.

\subsection{Participants and Apparatus}
The study procedure and materials received IRB approval.
We recruited 12 experienced XR developers and researchers (7 male, 5 female; aged 18–38) from a local university's VR clubs and relevant research groups. 
All participants reported normal or corrected-to-normal vision; 8 had over five years of AR/VR experience, and 9 rated their headset proficiency above 8 on a 10-point scale. 
Two participants joined pilot sessions that refined the study protocol, and thus were excluded from the ratings reported below.
Each participant received \$20 for a 75-minute session.

The study was conducted in a controlled indoor environment using an Apple Vision Pro, 
running a VisionOS~\cite{appleVisionOS} application developed with Unity's PolySpatial~\cite{unity_polyspatial_visionos}.
All sessions were audio- and video-recorded (including the participants' first-person views) for post-study analysis.

\subsection{Study Procedure}
Each session included the following phases:

\subsubsection{Introduction and Training (20mins)}
\noindent
Participants first signed the consent form and were briefed on the study's motivation, protocol, and \CoolName{} concepts. 
They then received a tutorial on the Apple Vision Pro's basic interactions (i.e., Gaze+Pinch). After calibrating eye–hand tracking, the experimenter demonstrated \CoolName{}
using a simplified office desktop scenario example.
Participants were encouraged to ask questions at any time, 
and the training phase continued until they felt confident performing the tasks on their own.

\vspace{3mm}
\begin{figure}[h]
  \centering
  \includegraphics[width=\columnwidth]{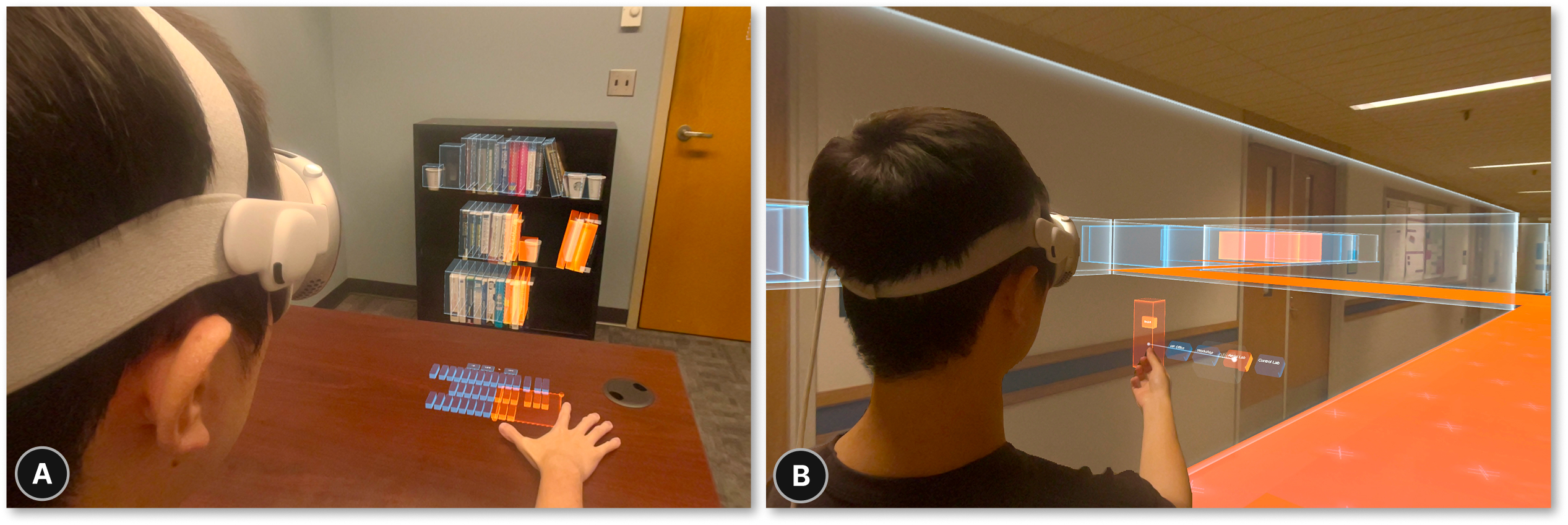}
  \caption{Two study tasks adapted from the applications.}
  \label{fig:study_tasks}
\end{figure}

\subsubsection{Tasks (40mins)}
\noindent
We designed two tasks (\autoref{fig:study_tasks}) to help participants experience the interaction features of \CoolName{}, 
each lasting about 20mins:
\begin{itemize}
    \item[\emph{Task}] 1 - \emph{\underline{Office Selection}}: 
    Adapted from the office scenario (Sec.\ref{sec:office}), where participants were asked to skim sticky notes and, based on the information provided, select books from a bookshelf and choose items from a rack within a given budget.

    \item[\emph{Task}]2 - \emph{\underline{Indoor Navigation}}: 
    Adapted from the building scenario (Sec.\ref{sec:building}), where participants were tasked with exploring rooms and selecting two specific rooms: Room A (
    a room on the next hallway wall pointed by the experimenter) 
    and Room B (the ``Aerial Lab'' in the Robotics department).
\end{itemize}

For both tasks, participants used the interaction features of \CoolName{} described in Sec.~\ref{sec:feats}.
To help contextualize the proxy-based approach,
they also had the option to further attempt the tasks using the default Gaze+Pinch method—upon selecting an object, a pop-up menu allowed object inspection and multi-object selection. 
Our goal was not to compare \CoolName{} against Gaze+Pinch—since the proxy system builds on top of it—but rather to ground expert feedback in a familiar interaction paradigm.
Participants were also given the opportunity to freely explore \CoolName{} features after completing the tasks.

\subsubsection{Semi-Structured Interview and Questionnaire (15mins)}
\noindent
At the end of the session, 
participants completed a follow-up questionnaire assessing the overall usefulness and usability of specific \CoolName{} interaction features using 7-point Likert items \add{(see the Appendix for the full list of questions)}, similar to other prior works~\cite{bovo2024embardiment}.
Then, the experimenter conducted a semi-structured interview to
further gather qualitative insights regarding proxy-based interaction's 
perceived benefits, 
limitations and areas for improvement,
and potential application scenarios.

\subsection{Expert Feedback}
\begin{figure*}[ht]
  \centering
  \includegraphics[width=1\textwidth]{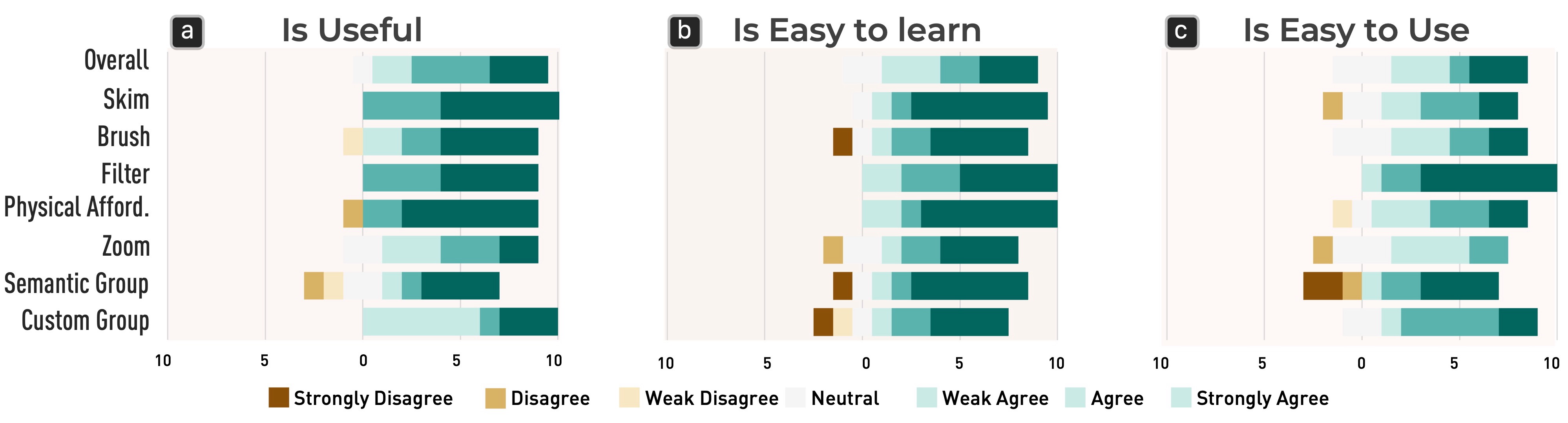}
  \caption{User feedback on various features, divided into three categories: usefulness, ease of learning, and ease of use. Each feature is rated on a scale from strongly disagree to strongly agree.}
  \label{fig:userfeedback}
\end{figure*}

All participants completed both tasks using \CoolName{}.  
\autoref{fig:userfeedback} shows that participants generally gave highly positive feedback on the overall system and individual features regarding usefulness, ease of learning, and ease of use.
\add{We performed a one-tailed single-sample Wilcoxon signed-rank test for each feature against the neutral midpoint of 4. With the exception of the Zoom ($p=.059$) and Reconfigure ($p=.08$) features in the ease of use category, all other features showed statistically significant positive ratings ($p<.05$) on usefulness, ease of learning, and ease of use.}
To better understand the reasons behind this feedback and identify areas for improvement, we analyzed the interview data using reflexive thematic analysis~\cite{braun2019reflecting}. We generated codes and themes both inductively (bottom-up) and deductively (top-down), focusing on participants' perceptions of strengths, limitations, and potential applications of the system.

Three authors collaboratively conducted the analysis, grouping codes into higher-level categories and focusing on participants' interaction behaviors. 
Disagreements were resolved through re-examination and discussion. 
We arrived at the final themes after more than three iterative rounds of discussion and refinement.

\subsubsection{Utility: Proxies Unlock Expressive and Effortless Interaction}

\noindent
Overall, participants consistently emphasized that \CoolName{}{} was \quot{very useful} (\autoref{fig:userfeedback}a, 9/10 rated positively) for interacting with real-world objects that might be physically distant, densely arranged, or otherwise inconvenient to access (e.g., E1, E4, E8). 
They praised the system's speed and convenience compared to conventional raycasting interactions, 
noting that \CoolName{} fluidly \quot{brings remote items closer} 
and simplifies complex selection tasks.
By decoupling interaction from physical constraints, \CoolName{} introduced several features that experts found particularly beneficial, such as reducing physical fatigue, offering richer, more expressive ways to convey interaction intent with objects, and facilitating forming a clear interaction intention of the scene objects.

\para{Reduce Physical Fatigue and Increase Accessibility.} 
A key benefit repeated across participants was how \CoolName{} reduces the physical overhead of interacting with distant or awkwardly placed objects. 
For instance, E1 explicitly mentioned that \quot{arms can get sore} from constantly pointing in VR and highlighted how \CoolName{} lets users \quot{bring it [the object] closer}. 
Similarly, E4 described the system as offering a \quot{miniature} in which users can use it 
reducing strain and making frequent object selections \quot{very helpful}.
Other participants pointed to the broader accessibility implications:
E10 pointed out that this proxy-based method can be especially valuable for users with limited mobility, while E7—who has ADHD—discussed how \quot{eye dancing} can interfere with constant gaze-based interactions. Having a proxy was deemed more accessible for users who frequently shift their eye focus.

\para{Increase Interaction Expressiveness Through Hierarchical and Semantic Structures.} 
Beyond mitigating fatigue, 
participants consistently highlighted how the proxy broadens interaction possibilities
by capturing hierarchical and semantics structures, 
rather than merely offering a digital copy of objects.
These additional layers enable users to express more nuanced intentions, rather than being limited to atomic object interactions. 
For instance, 
participants (e.g., E4, E5) appreciated the brush and container metaphor for 
collecting multiple items at once, fostering the expressions of complex intention.
E6 noted that the hierarchical views better match people's mental models of large or complex spaces, allowing them to select spatial objects more efficiently.
Likewise, 
E1 and E2 provided that the semantic attributes allow them to quickly interact with a subset of objects that would otherwise be tedious to handle.

In other words, it shortens the \emph{Articulatory Distance}~\cite{hitt1998difficulties} between what users want to express (e.g., multi-select a subset of items by attributes) and the physical gestures needed to accomplish it. 

\para{Enhance Users' Scene Understanding and Intent Formation via Structures.} 
Interestingly, participants also stressed that the AI-extracted structures in \CoolName{} 
improved their \emph{understanding} of real-world object organization, 
thereby easing the formation of interaction intention.
When interacting with real-world objects, users must first comprehend the scene, 
which inevitably demands cognitive resources. 
Multiple participants noted that the structured proxies in \CoolName{} reduce users’ cognitive load by removing the need to mentally organize objects themselves (E5), 
clarifying how objects relate to one another (E1, E3), 
and enhancing spatial awareness (E2) with simplified details (E10)
.
As a result, participants (e.g., E5, E9, E10) reported that these embedded structures supported smoother, more fluid interactions with real-world objects.

\subsubsection{Usability: Easy to Learn, but Opportunities for Refinement}
\noindent
Participants generally found \CoolName{} easy to learn (8/10 positive) and use (7/10 positive), as well as the individual interaction features (\autoref{fig:userfeedback}a and b).
Many emphasized that \CoolName{} \quot{feels straightforward} (E1) and \quot{doesn't need a lot of memorization} (E5), due in part to its reliance on familiar gestures. 
Moreover, the proxy representation enabled participants to understand interactions at a glance: \quot{when I look at the proxy, I know exactly ... what I'm supposed to do} (E1). 
Several (3/10) pointed to prior AR/VR experience as a factor facilitating quick adoption: \quot{I do research in VR … each single gesture is intuitive, and the operations inside [are] simple} (E3).
Nonetheless, participants also identified design improvements and potential usability challenges.
Below, we discuss two key themes that emerged from expert interviews.

\para{Learning Curve for New vs. Experienced Users.}
While participants emphasized \CoolName{}’s intuitive design, they differentiated between experienced XR users and newcomers. 
Many stressed that novices without extensive VR or gaming experience might require an initial acclimation period.
E7 estimated that a first‐time user could require 20 minutes (or more) to grasp the advanced features, such as groupings.
E8, a VR researcher, recalled teaching doctors the most basic controller interactions, 
suggesting that those lacking \emph{XR literacy} might need an explicit tutorial before feeling comfortable.
Nonetheless, the participants reasoned that after repeated exposure—\quot{a couple hours of using it} (E7)—users generally acquired enough comfort for daily tasks.

\para{Accuracy and Alignment Challenges.} 
Participants did report occasional misalignment and precision issues when interacting with proxies. 
For instance,
E4 observed some tracking errors \quot{…[brushing] boxes are not 100\% covering or at the place of [the proxies].} 
E5 and E7 mentioned difficulties in selecting the correct filter or proxy, 
often having to look down at the proxies rather than interacting with the real-world object directly. 
These issues were attributed to current implementation limitations, 
particularly in the proxy layout generation method (Sec.\ref{sec:generating}) and triggering mechanism (Sec.\ref{sec:interacting_proxies}). 
We consider this an inherent challenge in the system's current implementation design and 
had discussed potential improvements in the corresponding sections.

\subsubsection{Applicability: Across Diverse MR Scenarios}
\noindent
Beyond the motivating scenarios of selecting distant, crowded, or semi-occluded objects, 
participants identified wide-ranging scenarios in which \CoolName{} could be applied:

\para{Scenarios with Large-Scale Spatial Environments.}
Eight participants emphasized \CoolName{}'s benefits in spatial tasks in large-scale environments, 
such as navigating libraries, multi-floor buildings, or public venues. 
They noted that by abstracting physical layouts into abstract representations, 
users could rapidly locate items without repeatedly walking around. 
Some mentioned pathfinding (E2, E9), where \CoolName{} helps users gain instant overviews of complex spaces and easily identify targeted areas.

\para{Scenarios with Very Small or Hard-to-Reach Objects.}
Participants also highlighted \CoolName{}'s potential for selecting objects that are tiny, delicate, or physically inaccessible. 
E3, for example, envisioned using it to interact with objects in micro-scale scenarios (e.g., at the nano or cellular level), using a biomedical context as an example.
E10 foresaw medical applications, such as interacting with internal organs or areas one cannot physically touch. In these contexts, bringing minuscule or remote items into a manipulable proxy reduces errors and improves efficiency.

\para{Scenarios Requiring Accessibility.}
Several participants stressed how \CoolName{} might address accessibility challenges. 
E7, who experiences attention difficulties, explained that traditional gaze-based systems can become \quot{confused} when a user's eyes move unpredictably. 
By contrast, \CoolName{}'s proxy-based interaction accommodates intermittent or unreliable gaze. 
E10 likewise pointed out how individuals with limited mobility could benefit: \quot{If you don't have the ability to walk around a lot.}
These examples show \CoolName{}'s promise for users who need alternatives to decouple from the physical constraints.

\para{Scenarios Required Collaboration and Shared Understanding.}
E7, E10 suggested that proxies can foster collaborative workflows because they provide a persistent representation that can serve as a record or anchor for interactions. 
Users can save, share, and restore these proxies either asynchronously or among multiple people working in the same real-world environment. This level of continuity is difficult to achieve with conventional transient interactions such as raycasting.

\para{Scenarios for Selecting Physically Non-Existent Objects.}
Finally, \CoolName{} enables selection of virtual groupings or conceptual \quot{objects} that do not physically exist—providing an organizational layer beyond the real-world scene. 
E2 described how one might create a proxy group to classify related items, even though \quot{the organizer doesn't really exist.} 
This feature facilitates tasks such as conceptual planning, abstract organization, or advanced data visualization in both AR and VR settings.

\subsection{Observations and Suggestions}
We observed interesting behavior in the study which, together with the participants' suggestions, implies future design considerations.

\vspace{1mm}
\para{Proxies Foster Exploration of Real-World Objects.} 
We observed that, when using proxies to interact with real-world spaces, participants (e.g., E3, E4) often explored areas beyond their assigned tasks—for instance, visiting rooms in unrelated departments—and performed zoom-in/zoom-out operations to examine locations at different levels of detail. One participant (E8) even attempted to zoom in on a book to see its contents, despite not receiving formal instructions on the Zoom feature. 
These behaviors imply that our proxies offer a unique affordance or cognitive motivation~\cite{chen2013multimodal} to nudge proactive exploration of real-world environments. 
This could be especially beneficial for tasks in which thorough examination of physical surroundings are crucial, such as emergency rescue or other high-stakes scenarios. 
Future work can examine the differences in how various interface designs encourage (or discourage) user exploration, as well as the underlying reasons.

\para{Proxies May Decouple Users' Tasks from the Physical Scene.} 
Conceptually, proxies share a similar spirit with prior VR work such as SpaceTime~\cite{xia2018spacetime} and Poros~\cite{pohl2021poros}, in which a spatial or temporal context of the scene is captured and preserved, allowing users to interact with it afterwards. 
As a result, users might focus solely on the captured proxy, 
even though \CoolName{} was originally intended to keep them engaged with the actual, real-world environment.
For example, once E1 became comfortable with the system in Task 1, they mainly relied on proxy views rather than observing the physical scene.
This effect can be a double-edged sword, depending on the requirements of the task. 
If maintaining high situational awareness and focusing on the real environment is critical, 
users' declouping could become problematic. 
We highlight this observation to motivate future consideration on how to integrate proxies in ways that balance efficient task completion with necessary real-world engagement.

\para{Suggestions for Improvements.} 
Participants noted several limitations tied to system maturity and engineering robustness. 
For example, E2 and E4 reported unintentional actions caused by the gesture-detection limitations of AVP. E1, E3, and E7 found that the ``lazy follow'' method sometimes malfunctioned, leading to incorrect proxy placement. 
Several participants (e.g., E2, E4, E6) also highlighted suboptimal gesture designs among the introduced interaction features (Sec.~\ref{sec:feats}). 
E7 expressed concern about a lack of clear visual feedback when objects were added to a group, creating uncertainty during tasks. 
Indeed, these features were intended primarily for exploration and inspiration, so further refinement is warranted.
Beyond these usability gaps, 
 participants also raised broader issues regarding the reliability of AI-enabled interactions, integration with existing XR interaction ecosystems, and scaling to more complex scenes—topics we address in the Discussion section.

\section{Discussion}
\label{sec:discussion}

In this section, we reflect on our findings, 
outline future research directions, 
and acknowledge limitations in our study.

\para{Integrating Proxy Interactions with Mainstream MR Systems.}
One question that emerged during our expert interviews was how and when proxy-based interactions might coexist with system-level methods (e.g., raycasting) on mainstream MR devices (e.g., Quest 3 or AVP). 
Participants' opinions fell into two main categories. 
First, many felt that some basic proxy interactions—such as \emph{Skim} and \emph{Brush}—
could be integrated as OS-level features to extend or refine current default interaction techniques.
Second, participants were more cautious about adding advanced features—such as \emph{Zoom} or \emph{Filter}—as default interactions in the operating system. 
Instead, they recommended offering these functionalities in specific applications where nuanced user needs justify them.
Moreover, participants noted that once proxies are generated, many existing XR interaction methods could be adapted to improve accuracy and efficiency (e.g., disambiguation techniques to improve interaction accuracy).
Finally, they emphasized that if system-level proxy interactions rely on scene understanding, 
the underlying models must be sufficiently robust to handle varied real-world environments,
which we discuss further in the following section.

\para{Sustaining Human Agency in AI-Enabled Interactions.}
A critical consideration is whether the proxy reliably captures the user's intended object, i.e., it should include all relevant objects while excluding irrelevant ones. 
In scenarios such as the Building (Sec.~\ref{sec:building}) or Drone (Sec.~\ref{sec:drone}) examples—where proxies are derived from carefully modeled or sensor-tracked data—the resulting proxies tend to be highly accurate. 
By contrast, in scenarios that rely on AI scene parsing (e.g., the Office or Kitchen in Sec.~\ref{sec:office}), 
the AI may misclassify or overlook objects. 
Participants stressed that while improved AI models can mitigate such errors, 
a more robust approach is to allow humans step in and refine results.
For instance, if an object is missed, the user should be able to manually indicate its presence through traditional raycasting or bare-hand selections.
As highlighted in prior work~\cite{bovo2025symbiotic}, striking a balance between human oversight and AI automation is essential to achieving safe and reliable AI-enabled interactions in MR.
We therefore envision future proxy-based systems that treat AI and human inputs as complementary, letting users override or refine proxy generation on the fly, potentially with in-context learning~\cite{chulpongsatorn2023augmented}.

\para{Scaling to Complex, Dynamic Realities.}
Although our prototype demonstrates the feasibility of proxy-based interactions in relatively common scenarios, 
many real-world environments are far more complex. 
Unlike the mostly convex objects in our examples, real objects can exhibit irregular shapes, be non-rigid, or become tangled. This complexity complicates both scene understanding and interaction, as even well-established VR systems struggle with disambiguating these objects~\cite{dogan2024augmented}. 
Advanced disambiguation algorithms and better segmentation~\cite{tian2024diffuse} could be integrated into the proxy-interaction pipeline to handle such cases.
\add{Furthermore, to accommodate specific needs, we envision a customizable abstraction where designers or users can tune which physical properties (e.g., shape, scale, or position) are preserved in the proxy representation.}

An additional complication is scene dynamics: E5 and E7 raised the question of how proxies should adapt when real objects move in real time. 
\add{
While \CoolName{} currently targets static scenes, it can be extended to dynamic scenarios.
For example, in a basketball game, proxies can represent players tracked in real time using object tracking solutions (e.g., ByteTrack~\cite{zhang2022bytetrack}).
As the players move, the position of each proxy
would be to continuously updated.
Since movement is constrained to a fixed area (the court), users can easily select, filter, or group players to visualize in-game data—tasks that would be difficult with traditional raycasting methods. 
In other domains with larger or open areas (e.g., drone control), \CoolName{} remains useful as long as the object stays within visible range. If not, teleoperation may be more appropriate.
We envision that adapting \CoolName{} to dynamic scenes involves further design decisions, likely depending on the target application.
}

\para{Fostering Human–AI Collaboration through Interactions.}
Beyond basic selection and manipulation, proxies can foster human–AI collaboration by adding an interactive layer atop the physical environment.
Unlike traditional UIs in tightly controlled settings, the real world is both expansive and complex.
As observed in our study, AI can reveal scene structures (e.g., semantic clustering, spatial hierarchies) that assist the user's downstream tasks.
Yet the best way to present or interact with AI's scene understandings remains an open question~\cite{du2020depthlab}.
Prior MR+AI work~\cite{gunturu2024realitysummaryexploringondemandmixed} often annotates scenes with labels or visualizations,  
treating the user as a passive observer. 
By contrast, our approach turns these annotations into manipulable proxies, prompting the users to proactively \emph{explore} the AI's scene understanding,
akin to \emph{Explorable Explanations}~\cite{chulpongsatorn2023augmented}. 
We thus see designing such interactive, explodable proxy layers as a promising alternative 
to facilitate human-AI collaborations in MR.

\para{Limitations and Future Work.}
\add{Our study has several limitations that open avenues for future work.
First, one fundamental challenge of our design is the inherent trade-off introduced by abstraction. By modifying a representation's shape, scale, and layout, \CoolName{} may increase cognitive load, requiring users to constantly shift their attention between real-world targets and virtual proxies to maintain the mapping. While our design choices (e.g., finger-anchored placement and preserved relative layouts) aim to mitigate this cost, we acknowledge that this technique can make selection more difficult in some cases and that more work is needed to evaluate this balance.
Second,} our current system is a technology probe, which meant our primary goal was to implement and explore core proxy-based interactions rather than deliver a fully optimized solution.
As discussed in earlier sections, several technical details require further engineering iteration and testing, which is beyond the scope of this work.
Finally, following prior work~\cite{dogan2024augmented}, we conducted only qualitative evaluations with \add{a small example of} experienced XR users. While this approach yielded rich insights for iterative design\add{, it limits the generalizability of our findings to novice users. 
We did not conduct a direct quantitative comparison against a baseline, as no fair or appropriate baseline exists for comparison.
In the future, we plan to conduct dedicated follow-up studies to quantitatively evaluate user performance and experience. 
These studies will compare \CoolName{} with baseline techniques on relevant sub-tasks, incorporate standardized questionnaires such as the System Usability Scale, and involve a broader, more diverse user population across a wider range of scenarios.
}


\section{Conclusion}
We introduced Reality Proxy, a novel interaction approach for interacting with real-world objects in MR for digital tasks. 
Its key design principle is to shift interaction targets from physical objects to their abstract digital representations. 
By decoupling interaction from the physical constraints of real-world objects, Reality Proxy enables users to interact more effectively with objects that are crowded, distant, or partially occluded.
Augmented by AI, Reality Proxy further supports advanced MR interactions—such as multi-selection, semantic grouping, and spatial zooming—using intuitive direct manipulation gestures. Through several application scenarios, including everyday information retrieval, building navigation, and drone control, we demonstrated the versatility and effectiveness of this approach.
Our expert evaluation highlights Reality Proxy's potential to enhance user experience and interaction efficiency in MR environments, paving the way for more fluid, flexible, and expressive interaction with real-world objects in mixed reality.


\begin{acks}
The author(s) would like to express sincere gratitude to the anonymous reviewers for their constructive suggestions.
\end{acks}

\bibliographystyle{ACM-Reference-Format}
\bibliography{ref}

\add{
\appendix
\section*{APPENDIX}
\section{User Study Materials and Analysis}
\label{sec:appendix}

This appendix provides the questionnaires used in our expert evaluation and the statistical analysis of the post-task ratings.

\subsection{Post-Study Questionnaire}
For the post-study questionnaire, participants rated each of the 10 core features of \CoolName{} on a 7-point Likert scale based on the following three dimensions:
\begin{itemize}
    \item \textbf{Usefulness:} How useful was this feature for completing the tasks?
    \item \textbf{Ease of Learning:} How easy was it to learn how to use this feature?
    \item \textbf{Ease of Use:} How easy was this feature to use?
\end{itemize}
The features rated were: Skim, Brush, Filter, Physical Affordance, Zoom, Semantic Group, Custom Group, and an overall rating for the method.

\subsection{Post-Task Questionnaire and Analysis}

After each task, participants completed a questionnaire adapted from the System Usability Scale (SUS) assessing the overall system usability on a 7-point Likert scale (1 = Strongly Disagree, 7 = Strongly Agree). 
Note that we didn't collect feedback on each feature (where we did in the post-study) because not all features were applicable to every task (e.g., the brush or physical affordance were not needed in Task 2). 

\begin{itemize}
    \item I feel the interaction method is easy to learn in the task.
    \item I feel the interaction method is easy to use in the task.
    \item I feel this method can help me achieve my goal in the task efficiently.
    \item I feel satisfied with this interaction method in the task.
    \item I feel comfortable to use this interaction method in the task.
    \item I feel this method can help me achieve my goal in the task effectively.
    \item I feel in control with this method in the task.
    \item I would like to use this interaction method in my future MR device.
\end{itemize}

We analyzed the post-task ratings to assess the system's usability for each task and to check for differences between tasks:
\begin{itemize}[leftmargin=*]
    \item  \textbf{Comparison Between Task 1 and Task 2.}
A paired-sample Wilcoxon signed-rank test found no significant difference in ratings between Task 1 and Task 2 for any of the eight questions (all $p > .12$, all Cliff's $\delta < .05$), suggesting that participants' perceptions of the system were stable across both tasks.

\item \textbf{Ratings vs. Neutral.}
We conducted one-tailed single-sample Wilcoxon signed-rank tests to determine if the ratings for each question were significantly higher than the neutral midpoint of 4. As shown in \autoref{tab:posttask_analysis}, all questions received ratings that were significantly positive for both tasks. After applying a Holm-Bonferroni correction for multiple comparisons, all results remained significant. This indicates a strong positive reception of the system's usability in both task contexts.

\end{itemize}

We omitted these from the main text for space reasons and because the ratings were generally positive and aligned with our main findings, offering limited additional insight. 

\vspace{2mm}
\begin{table}[h]
\centering
\caption{Analysis of post-task ratings. One-tailed single-sample Wilcoxon signed-rank tests show all ratings were significantly greater than the neutral midpoint of 4.}
\label{tab:posttask_analysis}
\begin{tabular}{l|cc|cc}
\toprule
& \multicolumn{2}{c|}{\textbf{Task 1}} & \multicolumn{2}{c}{\textbf{Task 2}} \\
\textbf{Question} & \textbf{Z-value} & \textbf{p-value} & \textbf{Z-value} & \textbf{p-value} \\
\midrule
Easy to learn & 45.0 & .0020 & 44.0 & .0039 \\
Easy to use & 52.0 & .0059 & 53.0 & .0039 \\
Efficient & 53.5 & .0029 & 45.0 & .0020 \\
Satisfying & 45.0 & .0020 & 45.0 & .0020 \\
Comfortable & 36.0 & .0039 & 36.0 & .0039 \\
Effective & 53.5 & .0029 & 45.0 & .0020 \\
Controllable & 55.0 & .0010 & 36.0 & .0039 \\
Future intent & 45.0 & .0020 & 45.0 & .0020 \\
\bottomrule
\end{tabular}
\end{table}
}

\end{document}